\documentclass[prb, twocolumn, superscriptaddress, footinbib, amsmath, amsfonts, amssymb]{revtex4-2}

\usepackage[utf8]{inputenc}
\usepackage[T1]{fontenc}
\usepackage[main=english]{babel}

\usepackage{graphicx}
\usepackage[version=4]{mhchem}

\usepackage[scr=boondoxo, scrscaled=1.03]{mathalpha}

\usepackage{bm}
\usepackage{upgreek}

\usepackage{mleftright}
\usepackage{hyperref}
\hypersetup{colorlinks=true, allcolors=blue, breaklinks=true, pdfstartview=}

\makeatletter
\newcommand*{\defeq}{\mathrel{\rlap{%
\raisebox{0.3ex}{$\m@th\cdot$}}%
\raisebox{-0.3ex}{$\m@th\cdot$}}=}
\makeatother

\newcommand{\one}{\text{\usefont{U}{bbold}{m}{n}1}}
\MakeRobust{\one}

\newcommand*{\iu}{\mathrm{i}}
\newcommand*{\Elr}{\mathrm{e}}
\newcommand*{\Pauli}{\upsigma}
\newcommand*{\LCs}{\upepsilon}
\newcommand*{\Kd}{\updelta}
\DeclareMathOperator{\Dd}{\updelta}

\newcommand*{\abs}[1]{\mleft\lvert {#1} \mright\rvert}

\newcommand*{\dd}[2][]{\mathop{}\!\mathrm{d}^{#1} {#2}}
\newcommand*{\pdv}[2]{\frac{\partial #1}{\partial #2}}
\newcommand*{\var}[2][]{\mathop{}\!\delta_{#1} {#2}}

\newcommand*{\vdot}{\bm{\cdot}}
\newcommand*{\vcross}{\bm{\times}}
\newcommand*{\grad}{\bm{\nabla}}
\newcommand*{\vb}[1]{\bm{#1}}
\newcommand*{\vu}[1]{\bm{\hat{#1}}}

\newcommand*{\ket}[1]{\mleft\lvert {#1} \mright\rangle}
\newcommand*{\mel}[3]{\mleft\langle {#1} \middle| {#2} \middle| {#3} \mright\rangle}
\newcommand*{\ev}[1]{\mleft\langle {#1} \mright\rangle}

\newcommand*{\Hc}{\mathrm{H.c.}}
\DeclareMathOperator{\tr}{tr}
\DeclareMathOperator{\Tr}{Tr}
\DeclareMathOperator{\sgn}{sgn}

\newcommand*{\Z}{\mathbb{Z}}
\newcommand*{\R}{\mathbb{R}}
\let\C\relax
\newcommand*{\C}{\mathbb{C}}

\DeclareMathOperator{\Ugp}{U}
\DeclareMathOperator{\SU}{SU}
\DeclareMathOperator{\Ogp}{O}
\DeclareMathOperator{\SO}{SO}

\begin{document}
\title{Hopf symmetry protected topological phases in the vicinity of spin orders}
\author{Grgur Palle}
\email{gpalle@illinois.edu}
\affiliation{Department of Physics, The Grainger College of Engineering, University of Illinois Urbana-Champaign, Urbana, Illinois 61801, USA}
\affiliation{Anthony J.\ Leggett Institute for Condensed Matter Theory, The Grainger College of Engineering, University of Illinois Urbana-Champaign, Urbana, Illinois 61801, USA}
\affiliation{Institute for Theoretical Condensed Matter Physics, Karlsruhe Institute of Technology, 76131 Karlsruhe, Germany}
\date{\today}
\begin{abstract}
Hopf terms are topological theta terms that are associated with a host of interesting physics, including anyons, statistical transmutation, chiral edge states, and the spin quantum Hall effect.
Here, we show that Hopf terms can appear in two-dimensional metals without spin-orbit coupling in the vicinity of spin-ordered phases.
In their vicinity, their spin-like order parameters have a finite amplitude, but fluctuating orientation.
When both a magnetic and a spin loop-current order parameter fluctuate in the system, we show that the phase is governed by the Hopf term and realizes a Hopf symmetry protected topological phase.
This phase is protected by the unbroken $\SU(2)$ spin rotation symmetry, is gapped in the bulk, has chiral gapless edge states, and its spin-Hall conductance is quantized.
Lattice models that realize this phase are introduced.
In addition, we provide an elementary proof that the $\theta$ angle of the Hopf term must be quantized to multiples of $\pi$ in non-relativistic systems, thereby precluding anyonic skyrmions in condensed matter systems.
\end{abstract}

\maketitle

\section{Introduction}
Hopf terms are topological theta terms of 3-component unit-vector fields $\vu{n}(x)$ in 2+1D.
They were first introduced in an article by Wilczek and Zee~\cite{Wilczek1983} that showed that Hopf terms grant fractional statistics to skyrmions~\cite{Wilczek1983, Wu1984}, thus providing one of the first realizations of anyons~\cite{Leinaas1977, Goldin1980, Goldin1981, Wilczek1982, Wilczek1982-p2}.
Although it was soon realized that anyonic Hopf terms are not possible in non-relativistic systems~\cite{Polychronakos1987, Aitchison1989, Wen1991, Azcarraga1992, Baez1995, Bralic1996, Freed2018}, bosonic and fermionic Hopf terms have been found in many systems.
At first glance, the most frequently found bosonic Hopf terms seem uninteresting for they neither change the statistics of skyrmions nor add relative phases between topological sectors in the path integral.
Yet it is precisely for these same reasons that bosonic Hopf theta terms of the bulk can be recast into Wess-Zumino terms of the boundary.
When in the disordered phase, such Wess-Zumino terms are necessarily accompanied by chiral gapless edge states that carry quantized spin charges preferentially along one direction~\cite{Liu2013}.
Consequently, the spin-Hall response is quantized~\cite{Liu2013}.
Such phases that are gapped in the bulk and do not break symmetries, but are topologically non-trivial due to short-range entanglement, are called symmetry protected topological (SPT) phases~\cite{Gu2009, Senthil2015}.
When the Hopf term is fermionic, the disordered phase is either gapless and conformal, or gapped and twofold degenerate~\cite{Xu2013}.

\begin{figure}[t]
\includegraphics[width=\columnwidth]{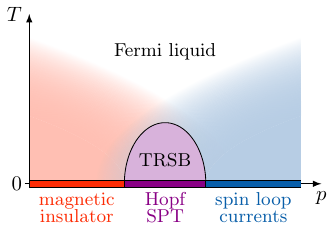}
\caption{Schematic temperature $T$ vs.\ tuning parameter $p$ phase diagram of a crossing between a magnetic and spin loop-current insulator in 2+1D.
Due to Hohenberg-Mermin-Wagner's theorem~\cite{Hohenberg1967, Mermin1966}, the magnetic (dark red) and spin loop-current (dark blue) phases, which break the $\SO(3)$ spin rotation symmetry, only set in at $T = 0$.
At finite $T$, the orientations of the corresponding order parameters $\vb{\phi}$ and $\vb{\phi}'$, respectively, fluctuate, as indicated by the light red ($\propto \abs{\vb{\phi}}^2$) and light blue ($\propto \abs{\vb{\phi}'}^2$) shading.
In between, the time-reversal-odd Ising variable $\vb{\phi} \vdot \vb{\phi}'$ may condense, resulting in a time-reversal symmetry-breaking (TRSB) phase (light purple).
Below it, if the spin rotation symmetry stays unbroken at $T = 0$, a Hopf symmetry protected topological (SPT) phase will generically appear (dark purple).}
\label{fig:phase-diagram}
\end{figure}

In light of the importance of the topological theta term in antiferromagnetic Heisenberg spin chains~\cite{Haldane1983, Haldane1983-p2, Haldane2016, Affleck1989}, many have searched for the Hopf term in two-dimensional antiferromagnetic Heisenberg models~\cite{Dzyaloshinskii1988}.
Unlike in the 1D case, the Hopf term is non-local in the original spin variables, and a total derivative in the gauge-redundant variables~\cite{Wu1984}, so finding it is a delicate affair.
All studies to date~\cite{Dombre1988, Fradkin1988, Haldane1988, Ioffe1988, Dombre1989} have found that there is no Hopf term, whether on a square or triangular lattice~\footnote{Some have expressed reservations regarding these results~\cite{Baez1991, Balakrishnan1993} and one study claims to have found a Hopf term~\cite{VillainGuillot1998}. Upon repeating their calculation~\cite{VillainGuillot1998}, however, one finds that in their Eq.~(25) the term $\partial_x \vb{\Gamma}_y - \partial_y \vb{\Gamma}_x$ should appear with a plus sign, thereby invalidating any relation between this term and the Hopf term. That a plus sign must appear here also follows from symmetries. From their unnumbered equation that precedes Eq.~(16), it follows that $\vb{\Gamma}_{\mu}$ and $\partial_{\mu} \vb{n}$ transform the same. Since $\vb{n}$ is antiferromagnetically modulated ($\vb{Q} = (\pi, \pi)$), $(\vb{\Gamma}_x, \vb{\Gamma}_y)$ transforms like $(y, x)$ and only the plus-sign combination $\partial_x \vb{\Gamma}_y + \partial_y \vb{\Gamma}_x$ transforms trivially, just like the ferromagnetically modulated $\vb{l}$ of their Eq.~(25). Explicit calculations confirm this. Surprisingly, in their earlier paper~\cite{VillainGuillot1996} they had the correct sign.}.

If one couples fermions to a 3-component unit-vector field $\vu{n}(x)$, their fluctuations may induce a Hopf term in the effective action of $\vu{n}$.
Within gradient expansions, such Hopf terms are a result of radiative (or loop) corrections and are called radiatively-induced~\cite{Dunne1999}.
Using gradient expansions, radiatively-induced Hopf terms have been found in thin films of spin-triplet superfluid $^3$He-A~\cite{Volovik1989}, $\nu = 1$ quantum Hall states that retain spin freedom due to weak Zeeman coupling~\cite{Moon1995, Iordanskii1997, Apel1997, Ray1999, Sengupta2000}, a modified Haldane's model~\cite{Yakovenko1990}, quasi-1D organic conductors with spin-density waves~\cite{Yakovenko1991}, and relativistic fermion models with Yukawa coupling to isospin~\cite{Hlousek1990}.
In all cases, the Hopf term coefficient $\theta$ is found to be $\pi$ times the Chern number $C$ that characterizes the fermionic band Hamiltonian with $\vu{n}(x)$ set to a constant~\footnote{Before the earliest gradient expansion study that found a Hopf term~\cite{Volovik1989}, the Hubbard model on a 2D square lattice was also studied using gradient expansions~\cite{Wen1988} and found to not possess a Hopf term, in retrospect because its Chern number vanishes. Since the Hubbard model at half-filling and strong-coupling maps onto the antiferromagnetic Heisenberg model, this null result complements other studies~\cite{Dombre1988, Fradkin1988, Haldane1988, Ioffe1988, Dombre1989} that found no Hopf term.}.
However, Hopf terms can arise even when the Chern number vanishes, as is the case for Dirac systems.
The simple form of the Dirac Hamiltonian enables the non-perturbative calculation of the Hopf term's $\theta$ angle, whether by embedding $\vu{n}$ in a larger manifold that has no topological invariants in 2+1D~\cite{Abanov2000, Abanov2001} or by calculating the phase difference for a specific field configuration with a finite Hopf number~\cite{Abanov2000-p2}.
Such Dirac models with Yukawa coupling to $\vu{n}$ have been used to describe topological superconductors~\cite{Abanov2001-p2}, superconducting boundaries of topological insulators~\cite{Ran2011}, and lattice models tuned close to Dirac points~\cite{Grover2008, Yue2023, Yue2025}.

In this article, we consider radiatively-induced Hopf terms near spin-ordered phases whose order parameters couple to electrons via Hund's coupling ($\sim \vb{\phi} \vdot \vb{\Pauli}$).
To retain the $\SO(3)$ spin symmetry that is necessary for the dynamical relevance of topological terms, we assume negligible spin-orbit coupling and focus on the region surrounding the spin-ordered phases where the spin order has not yet set in and where the spin rotation symmetry is unbroken.
We study the necessary conditions for the appearance of a Hopf term and find that simultaneous presence of a magnetic $\vb{\phi}$ and a spin loop-current $\vb{\phi}'$ order parameter is needed to break time-reversal (TR) symmetry without breaking the spin rotation symmetry.
We therefore consider the crossing of magnetic and spin loop-current order, as depicted in Fig.~\ref{fig:phase-diagram}.
In between, we show that a partially ordered phase can arise which breaks TR through the condensation of $\vb{\phi} \vdot \vb{\phi}'$, while retaining the spin rotation symmetry by not having $\vb{\phi}$ or $\vb{\phi}'$ condense.
Drawing on previous work~\cite{Liu2013, Yue2023}, we argue that this phase is an SPT phase with chiral gapless edge states and a quantized spin-Hall conductance.
A number of lattice models for this phase are introduced and studied.
We also discuss why the $\theta$ of the Hopf term is always quantized in lattice models appropriate for condensed matter physics, as this is not always respected in relativistic continuum (Dirac model) descriptions.

The paper is organized into two parts.
The first part (Sec.~\ref{sec:Hopf-term}) deals with the general definition and physical consequences of a Hopf term, while the second part (Sec.~\ref{sec:analysis}) derives a Hopf term in Hund-coupled spin-ordered insulators that have no spin-orbit coupling (Fig.~\ref{fig:phase-diagram}).
The first part starts with the definition of a Hopf term in the simplest case of a relativistic spacetime.
In Sec.~\ref{sec:theta-quant} we then explain why this definition needs to be modified when matter breaks relativistic covariance and how this implies that the Hopf $\theta$ is quantized, before finally turning to the physical consequences of a Hopf term in Sec.~\ref{sec:Hopf-properties}.
We begin the second part by introducing a general lattice model of a Hund-coupled spin-ordered insulator.
Radiatively-induced Hopf terms are calculated for this model in Sec.~\ref{sec:Hopf-gradient} and the necessary conditions for their appearance are derived in Sec.~\ref{sec:Hopf-symmetry} thereafter.
In Sec.~\ref{sec:lattice-model} we introduce concrete lattice models that realize the Hopf SPT.
We conclude with a discussion of experimental signatures and possible realization of our proposal, as well as open problems.

\section{The Hopf term} \label{sec:Hopf-term}
In this part, we define the Hopf term, show that it is quantized for non-relativistic spacetimes, and discuss its physical consequences.

\subsection{Definition and interpretation} \label{sec:Hopf-def}
The Hopf term~\cite{Wilczek1983, Wu1984, Azcarraga1992} is a topological theta term~\cite{Altland2010} constructed from the Hopf invariant $Q_{\text{Hopf}} \in \Z$, which is a topological invariant that enumerates the topologically distinct mappings from the 3-sphere $\mathcal{S}^3$ to the 2-sphere $\mathcal{S}^2$~\cite{Hopf1931, Naber2011}.
The corresponding homotopy group is $\pi_3(\mathcal{S}^2) = \Z$.
Here, we precisely define the Hopf term and build some physical intuition regard it, both of which shall be important later on.

In physical settings, mappings from $\mathcal{S}^3$ to $\mathcal{S}^2$ are realized through vector-valued fields $\vb{\phi}(x) = \bar{\phi} \, \vu{n}(x) \in \R^3$ whose amplitude $\abs{\vb{\phi}(x)} = \bar{\phi} \neq 0$ has condensed, but its direction $\vu{n}(x) \in \mathcal{S}^2$ still fluctuates.
Here $x \equiv x^{\mu} \equiv \vb{x} = (\tau, \vb{r})$ are the coordinates of the 2+1D spacetime.
In the systems that we shall study (Fig.~\ref{fig:phase-diagram}, Sec.~\ref{sec:analysis}), $\vb{\phi}$ is the order parameter of a magnetic or spin loop-current phase.
In fact, as reviewed in the introduction, most studies to date had such spin-like order parameters for $\vb{\phi}$ that either described magnetic phases~\cite{Dombre1988, Fradkin1988, Haldane1988, Ioffe1988, Dombre1989, Moon1995, Iordanskii1997, Apel1997, Ray1999, Sengupta2000, Yakovenko1990, Yakovenko1991, Yue2023, Yue2025} or triplet superfluids~\cite{Volovik1989, Abanov2001-p2, Ran2011}.
This is because an $\SO(3)$ symmetry is needed to enable the fluctuations of $\vu{n}$ at low energies, and the main way of realizing this symmetry is through spin, while having no spin-orbit coupling.

To study hopfionic instanton events, one compactifies the $\R^3$ spacetime to a 3-sphere $\mathcal{S}^3$, implicitly assuming that $\vu{n}(x)$ goes to a constant at spacetime infinity.
Notice how this compactification scheme presumes that space and time are interchangeable -- a point that we shall come back to in the next section.
The Hopf invariant is now given by the non-local expression
\begin{align}
Q_{\text{Hopf}}[\vu{n}] &= \int \dd[3]{x} \dd[3]{x'} \frac{(\vb{x} - \vb{x}') \vdot \mleft[\vb{f}(\vb{x}) \vcross \vb{f}(\vb{x}')\mright]}{4 \pi \abs{\vb{x} - \vb{x}'}^3}, \label{eq:Hopf-non-loc}
\end{align}
where the integral goes over the whole spacetime and $\vb{f}(x)$ is the topological field strength
\begin{align}
f^{\mu}(x) &\defeq \frac{1}{8 \pi} \LCs^{\mu \nu \rho} \vu{n}(x) \vdot \mleft[\partial_{\nu} \vu{n}(x) \vcross \partial_{\rho} \vu{n}(x)\mright]. \label{eq:topo-f-def}
\end{align}
Here and elsewhere summations over repeated $\mu, \nu, \rho \in \{0, 1, 2\}$ are implicit, we use Euclidean signature ($g_{\mu \nu} = g^{\mu \nu} = \Kd_{\mu \nu}$), and $\LCs^{\mu \nu \rho} = \LCs_{\mu \nu \rho}$ is the Levi-Civita symbol.

\begin{figure}[t]
\includegraphics[width=\columnwidth]{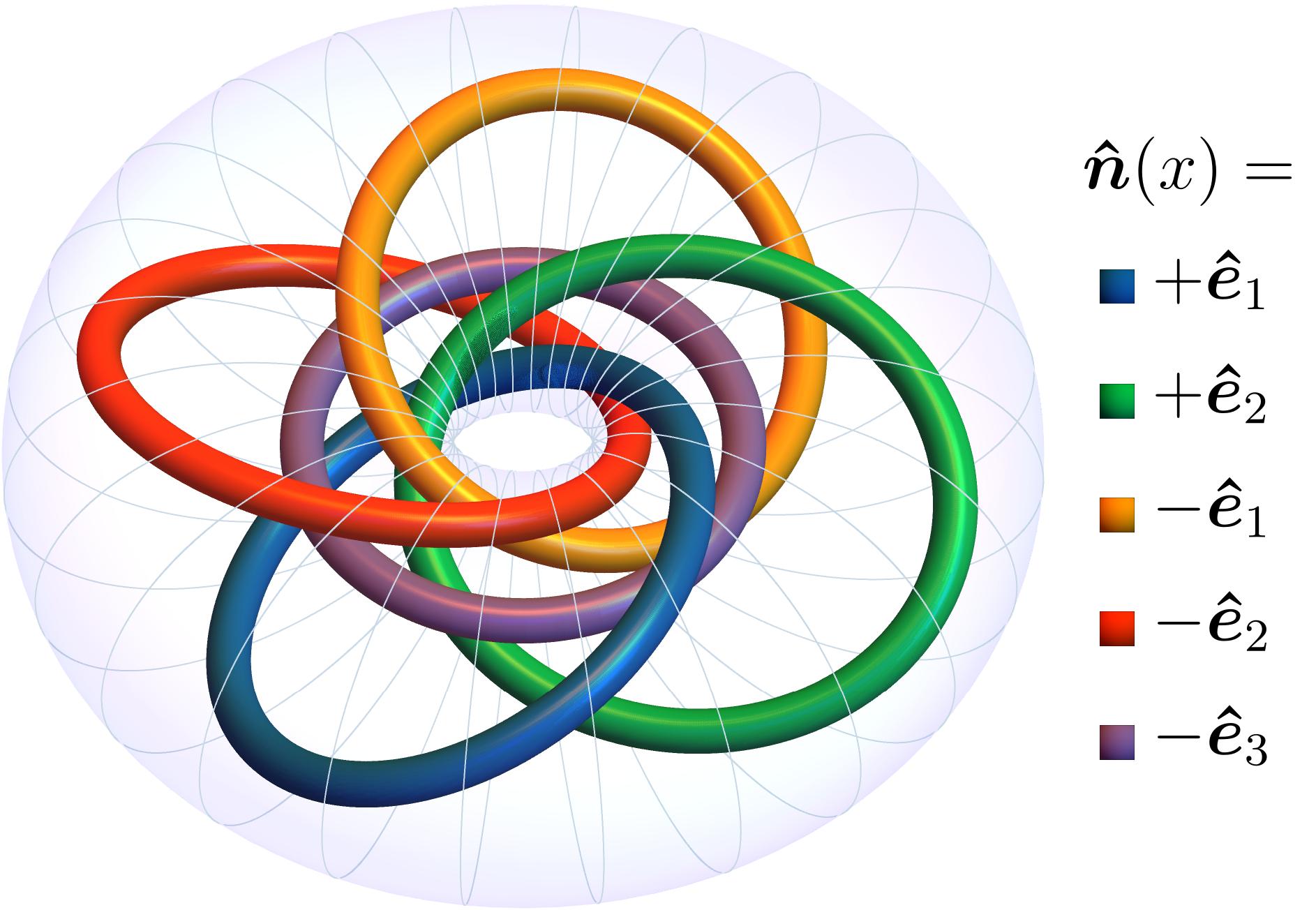}
\caption{An example of a field configuration $\vu{n}(x)$ that has a unit Hopf number $Q_{\text{Hopf}} = +1$.
Everywhere outside of the lightly blue-shaded torus, $\vu{n}(x)$ points along $+ \vu{e}_3$  to within an angle of $\pi / 3$.
Within the torus, the thick lines (tubes) denote regions where $\vu{n}(x)$ is oriented along the directions indicated by the legend.
One way of interpreting this field configuration is as the creation of a skyrmion--anti-skyrmion pair that then winds around itself before getting annihilated.}
\label{fig:hopfion}
\end{figure}

Intuitively, the Hopf invariant measures the way magnetic (or baby) skyrmions wind with themselves or around each other.
Skyrmions are, let us recall, solitons that are topologically protected by their winding number invariant, which is related to the homotopy group $\pi_2(\mathcal{S}^2) = \Z$~\cite{Fert2017, Petrovic2025}.
For a 2D surface $\Sigma$ without a boundary ($\partial \Sigma = 0$), the integer-valued skyrmion or winding number of $\mleft.\vu{n}(x)\mright|_{x \in \Sigma}$ restricted to $\Sigma$ equals
\begin{align}
Q_{\text{sky}} &= \int_{\Sigma} \dd{\vb{S}} \vdot \vb{f}(x). \label{eq:skyrmion}
\end{align}
If a skyrmion is oriented along $\vu{n}_c$ in its core, its trajectory is specified by the path $\gamma_{\ell}(\vu{n}_c)$ made of points $x_c \in \gamma_{\ell}(\vu{n}_c)$ such that $\vu{n}(x_c) = \vu{n}_c$.
Here the index $\ell$ enumerates the skyrmions.
More precisely, if two $\gamma_{\ell}$ are far apart we speak of distinct skyrmions, whereas a group of $Q_{\text{sky}}$ paths $\gamma_{\ell}$, $\ell \in \{1, \ldots, Q_{\text{sky}}\}$, moving together constitute one skyrmion of charge $Q_{\text{sky}}$.
Whether a path $\gamma_{\ell}$ corresponds to a skyrmion or anti-skyrmion depends on whether $f^0(x)$ is positive or negative for $x \in \gamma_{\ell}$, respectively.
As we prove in App.~\ref{sec:link-formula}, the Hopf invariant can be expressed in terms of linking numbers between these paths:
\begin{align}
Q_{\text{Hopf}} &= \sum_{\ell \ell'} \operatorname{link}\mleft[\gamma_{\ell}(\vu{n}_c), \gamma_{\ell'}(\vu{n}_c')\mright], \label{eq:Hopf-linking}
\end{align}
where $\vu{n}_c$ and $\vu{n}_c'$ are fixed, but different, unit vectors.
A field configuration with a finite Hopf invariant, often referred to as a hopfion, is illustrated in Fig.~\ref{fig:hopfion}.

The Hopf term is defined as~\cite{Wu1984}:
\begin{align}
S_{\text{Hopf}}[\vu{n}] &\defeq - \iu \, \theta \, Q_{\text{Hopf}}[\vu{n}], \label{eq:Hopf-th-term}
\end{align}
where $\theta \in \R$ is the angle that specifies this theta term, $Q_{\text{Hopf}}$ is integer-valued for $\vu{n}(x)$ that go to the same constant for all directions at infinity, and an imaginary unit appears explicitly because the action is Euclidean ($\Elr^{- S_{\text{Hopf}}[\vu{n}]} = \Elr^{\iu \theta Q_{\text{Hopf}}[\vu{n}]}$ is a phase, as it should be).

To see how the Hopf term affects skyrmion statistics, consider a skyrmion of charge $Q_{\text{sky}}$ that is made of $Q_{\text{sky}}$ paths $\gamma_{\ell}$, $\ell \in \{1, \ldots, Q_{\text{sky}}\}$.
From Eq.~\eqref{eq:Hopf-linking} it follows that the phase changes by $\Elr^{\iu \theta Q_{\text{sky}} Q_{\text{sky}}'}$ upon the exchange of two skyrmions of charge $Q_{\text{sky}}$ and $Q_{\text{sky}}' \in \Z$, but also that $2 \pi$ rotations of one skyrmion result in the phase factor $\Elr^{\iu \theta Q_{\text{sky}}^2}$.
Hence the Hopf term~\eqref{eq:Hopf-th-term} grants fractional spin $S = \frac{\theta}{2 \pi} Q_{\text{sky}}^2$ and statistics to skyrmions~\cite{Wilczek1983}.
When $\theta = 2 n \pi$ or $(2n+1) \pi$ for some $n \in \Z$, we shall call the Hopf term bosonic or fermionic, respectively.
We discuss further consequences of a Hopf term in Sec.~\ref{sec:Hopf-properties}.

\subsection{Quantization of theta} \label{sec:theta-quant}
The expression~\eqref{eq:Hopf-non-loc} for the Hopf invariant only applies when the skyrmion number $Q_{\text{sky}} = \int \dd{x^1} \dd{x^2} f^0(x)$ vanishes.
At finite temperatures, spacetime is periodic in imaginary time $\tau \equiv x^0$ and topological sectors with a finite number of skyrmions, although of higher energy than the $Q_{\text{sky}} = 0$ sector, are not completely suppressed by their Boltzmann factors.
Hence spacetime compactifies to $\mathcal{S}^1_{\tau} \times \mathcal{S}^2_{\vb{r}}$ rather than $\mathcal{S}^3_x$ and $\vu{n}(x)$ configurations with finite $Q_{\text{sky}}$ appear in the path integral.
Because of these two non-relativistic effects, $\theta$ must be quantized to multiples of $\pi$~\cite{Polychronakos1987, Aitchison1989, Wen1991, Azcarraga1992, Baez1995, Bralic1996, Freed2018}.
Anyonic skyrmions are therefore not possible in condensed matter systems.
Here, we prove this statement.
Although this result is important, the proof that follows is technical and can be skipped without affecting the understanding of subsequent sections.

To recast the Hopf invariant, notice that from $\vu{n}^2 = 1$ it follow that $\partial_{\mu} \vu{n}$ is orthogonal to $\vu{n}$.
Hence there are only two linearly independent $\partial_{\mu} \vu{n}$.
The divergence of the topological field strength $\partial_{\mu} f^{\mu} \propto \det\mleft(\partial_{\mu} \vu{n}, \partial_{\nu} \vu{n}, \partial_{\rho} \vu{n}\mright)$ therefore vanishes identically:
\begin{align}
\grad \vdot \vb{f}(x) = 0.
\end{align}
Locally, we may thus always write it in terms of a vector potential:
\begin{align}
\vb{f}(x) = \grad \vcross \vb{a}(x). \label{eq:vec-pot-a}
\end{align}
Globally, however, this cannot be done whenever $Q_{\text{sky}}$ is non-zero because a global $\vb{a}(x)$ would imply that
\begin{align}
Q_{\text{sky}} &= \int_{\mathcal{S}^2_{\vb{r}}} \dd{\vb{S}} \vdot \mleft(\grad \vcross \vb{a}\mright) = \oint_{\partial \mathcal{S}^2_{\vb{r}}} \dd{\vb{\ell}} \vdot \vb{a} = 0,
\end{align}
as follows from Stokes' theorem and the compactification of space ($\partial \mathcal{S}^2_{\vb{r}} = 0$).
Note that $\dd{\vb{S}} = \vu{e}_0 \dd{x^1} \dd{x^2}$ for $\mathcal{S}^2_{\vb{r}}$.
Said differently, $\vb{a}(x)$ asymptotically decays like $Q_{\text{sky}} / r$ because of the finite ``flux'' through the $\tau = 0$ plane.
This, in turn, obstructs the introduction of an $\vb{a}(x)$ that is well-defined at compactified spatial infinity, reflecting the presence of a Dirac string at $r \to + \infty$.

In the $Q_{\text{sky}} = 0$ case, the global $\vb{a}(x)$ can be used to recast Eq.~\eqref{eq:Hopf-non-loc} into a local form:
\begin{align}
Q_{\text{Hopf}} &= \int \dd[3]{x} \, \vb{a} \vdot \vb{f} = \int \dd[3]{x} \, \LCs^{\mu \nu \rho} a_{\mu} \partial_{\nu} a_{\rho}. \label{eq:Hopf-CS}
\end{align}
This is possible at the expense of introducing a gauge redundancy $\vb{a} \mapsto \vb{a} + \grad \chi$.
Eq.~\eqref{eq:Hopf-non-loc} is recovered by solving $\vb{f} = \grad \vcross \vb{a}$ in the Coulomb gauge $\grad \vdot \vb{a} = 0$.
In terms of $a_{\mu}(x)$, we thus see that the Hopf term~\eqref{eq:Hopf-th-term} takes the form of an abelian $\Ugp(1)$ Chern-Simons term~\cite{Chern1974, Baez1994}:
\begin{align}
S_{\text{CS}}[a] &= - \iu \, k \, \pi \int \dd[3]{x} \, \LCs^{\mu \nu \rho} a_{\mu} \partial_{\nu} a_{\rho}.
\end{align}
The Chern-Simons level $k$ is an integer whenever large gauge transformations exist~\cite{Baez1994}.
Since large gauge transformations exist for $\mathcal{S}^1_{\tau} \times \mathcal{S}^2_{\vb{r}}$, namely those where $\Ugp(1)$ winds in imaginary time, it follows that $\theta = k \pi$ is quantized~\cite{Polychronakos1987, Aitchison1989, Wen1991, Azcarraga1992, Baez1995, Bralic1996, Freed2018}.
An example of a large gauge transformation is $\chi(x) = 2 \pi m \, x^0 / \beta$, where $m \in \Z$ is the winding number.

It is instructive to see a direct proof for why $\theta$ must be quantized.
The currently available proofs~\cite{Polychronakos1987, Aitchison1989, Wen1991, Azcarraga1992, Baez1995, Bralic1996, Freed2018} all employ a Čech cohomology construction~\cite{Polychronakos1987, Alvarez1985} that is not easily accessible to most practicing physicists.
Here we provide an elementary proof.

\begin{figure}[t]
\includegraphics[width=\columnwidth]{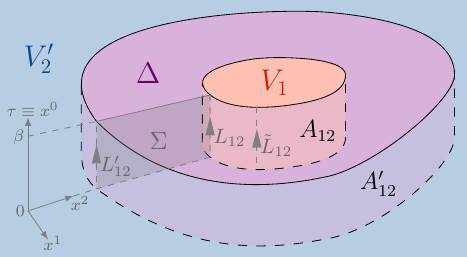}
\caption{Partition of compactified spacetime $\mathcal{S}^1_{\tau} \times \mathcal{S}^2_{\vb{r}}$ into $V_1 = V_1' - \Delta$ (red region) and $V_2 = V_2' + \Delta$ (blue and purple regions).
Since spacetime is periodic in imaginary time $\tau \equiv x^0$, which corresponds to height in the figure, the drawn cylinders represent toruses.
Here $\Delta$ is the volume change under the deformation $V_n \to V_n'$.
$A_{12} = \partial V_1 = - \partial V_2$, $A_{12}' = \partial V_1' = - \partial V_2'$, and $\Sigma$ are 2D surfaces, whereas $L_{12}$, $L_{12}'$, and $\tilde{L}_{12}$ are 1D lines.
$\partial \Delta = A_{12}' - A_{12}$ and $\partial \Sigma = L_{12} - L_{12}'$.}
\label{fig:patches}
\end{figure}

We start by generalizing the expression~\eqref{eq:Hopf-CS} for the Hopf invariant to the case of non-zero $Q_{\text{sky}}$.
Naively, one might attempt to simply define it piecewise like so:
\begin{align}
\int_{V_1} \dd[3]{x} \, \vb{a}_1 \vdot \vb{f} + \int_{V_2} \dd[3]{x} \, \vb{a}_2 \vdot \vb{f},
\end{align}
where $V_1 \cup V_2 = \mathcal{S}^1_{\tau} \times \mathcal{S}^2_{\vb{r}}$ partition spacetime (Fig.~\ref{fig:patches}) and $\grad \vcross \vb{a}_n = \vb{f}$ on each $V_n$.
However, this expression depends on both the gauge $\vb{a}_n \mapsto \vb{a}_n + \grad \chi_n$ and the $V_n \to V_n'$ used to partition spacetime.

To introduce the necessary correction terms, first note that by Stokes' theorem
\begin{align}
Q_{\text{sky}} &= \mleft.\oint \dd{\vb{\ell}} \vdot \mleft(\vb{a}_1 - \vb{a}_2\mright)\mright|_{x^0 = \tau},
\end{align}
where the loop integral goes over a fixed-$\tau$ cross-section of $A_{12} = \partial V_1 = - \partial V_2$.
We can therefore write $\vb{a}_1 - \vb{a}_2 = \grad \psi_{12}$ only if we allow $\psi_{12}$ to have a jump of $Q_{\text{sky}}$ at a certain $\vb{r}_{\tau}$ for all $\tau$.
These points of discontinuity constitute a time-like curve $L_{12} = \mleft\{(\tau, \vb{r}_{\tau}) \mid \tau \in [0, \beta]\mright\}$ contained in $A_{12}$, as shown in Fig.~\ref{fig:patches}.
The Hopf invariant is now given by:
\begin{align}
\begin{aligned}
Q_{\text{Hopf}} &= \int_{V_1} \dd[3]{x} \, \vb{a}_1 \vdot \vb{f} + \int_{V_2} \dd[3]{x} \, \vb{a}_2 \vdot \vb{f} \\
&\quad - \int_{A_{12}} \dd{\vb{S}} \vdot \vb{f} \, \psi_{12} + Q_{\text{sky}} \int_{L_{12}} \dd{\vb{\ell}} \vdot \vb{a}_1.
\end{aligned}  \label{eq:Hopf-CS-v2}
\end{align}
This expression is invariant under local gauge transformations $\vb{a}_n \mapsto \vb{a}_n + \grad \chi_n$ because $\psi_{12} \mapsto \psi_{12} + \chi_1 - \chi_2$ simultaneously changes to cancel any gauge-dependent terms.
Due to the last line integral, it is also invariant if we move the jump line $L_{12} \to \tilde{L}_{12}$ by increasing $\psi_{12}$ by $Q_{\text{sky}}$ on a $\tilde{\Sigma}$ contained in $A_{12}$ whose $\partial \tilde{\Sigma} = L_{12} - \tilde{L}_{12}$.
Increasing $\psi_{12} \mapsto \psi_{12} + c$ globally by a constant also doesn't affect $Q_{\text{Hopf}}$ because $\int_{A_{12}} \dd{\vb{S}} \vdot \vb{f} = \oint_{\partial A_{12}} \dd{\vb{\ell}} \vdot \vb{a}_1 = 0$, as follows from $\partial A_{12} = 0$.
Finally, Eq.~\eqref{eq:Hopf-CS-v2} is invariant under deforming $V_1 \to V_1' = V_1 + \Delta$ and $V_2 \to V_2' = V_2 - \Delta$ as depicted in Fig.~\ref{fig:patches}, as one can show by exploiting the relations $\partial \Delta = A_{12}' - A_{12}$, $\vb{a}_1(x) - \vb{a}_2(x) = \grad \psi_{12}(x) + Q_{\text{sky}} \int_{\Sigma} \dd{\vb{S}_{x'}'} \Dd(x-x')$, and $\partial \Sigma = L_{12} - L_{12}'$.

Under a large gauge transformation $\vb{a}_n \mapsto \vb{a}_n + \grad \chi$ where $\chi(x) = 2 \pi m \, x^0 / \beta$ with $m \in \Z$, we find that
\begin{align}
Q_{\text{Hopf}} &\mapsto Q_{\text{Hopf}} + 2 m Q_{\text{sky}},
\end{align}
where the additional factor of two arises because of the $L_{12}$ line integral in Eq.~\eqref{eq:Hopf-CS-v2}~\cite{Polychronakos1987}.
Hence to ensure the gauge invariance of the topological Hopf phase factor
\begin{align}
\Elr^{- S_{\text{Hopf}}} = \Elr^{\iu \theta Q_{\text{Hopf}}} \mapsto \Elr^{\iu \theta 2 m Q_{\text{sky}}} \Elr^{- S_{\text{Hopf}}},
\end{align}
$\theta = k \pi$ must be quantized, i.e., $k \in \Z$.

For compactified relativistic spacetimes $\cong \mathcal{S}^3_x$, $Q_{\text{sky}}$ vanishes identically.
This holds because space contracts to a point at temporal infinity, implying $Q_{\text{sky}} = 0$ as $\tau \to \pm \infty$.
Topological continuity then enforces $Q_{\text{sky}} = 0$ for all time.
This is why anyonic $\theta$ are possible in relativistic theories over $\mathcal{S}^3_x$ (but not more generally~\cite{Freed2018}).

\subsection{Physical implications} \label{sec:Hopf-properties}
Here, we discuss the physical implications of a Hopf term~\cite{Wilczek1983, Liu2013}.
Many of these same points were discussed recently in Ref.~\cite{Yue2023}.
Throughout we shall assume that $\vu{n}(x)$ is a spin-like TR-odd field that gaps out the itinerant fermions via Hund's coupling ($\sim \vu{n} \vdot \vb{\Pauli}$), as in the model of the next part (Sec.~\ref{sec:general-model}).

One much-discussed implication of a Hopf term is that it changes the spin and statistics of skyrmions~\cite{Wilczek1983}.
As explained in Sec.~\ref{sec:Hopf-def}, a Hopf term induces fractional spin
\begin{align}
S &= \frac{\theta}{2 \pi} Q_{\text{sky}}^2
\end{align}
to skyrmions of charge $Q_{\text{sky}}$ and modifies their statistics so that under exchange the phase changes by $\Elr^{\iu \theta Q_{\text{sky}}^2}$.
More generally, under the exchange of two skyrmions of charge $Q_{\text{sky}}$ and $Q_{\text{sky}}'$ the phase changes by $\Elr^{\iu \theta Q_{\text{sky}} Q_{\text{sky}}'}$.
However, as we proved in the last section, $\theta$ is quantized in all condensed matter systems~\cite{Polychronakos1987, Aitchison1989, Wen1991, Azcarraga1992, Baez1995, Bralic1996, Freed2018}, thereby precluding anyonic skyrmions.
For the intersection of magnetic and spin loop-current order that is the focus of our article (Fig.~\ref{fig:phase-diagram}), we find that
\begin{align}
\theta &= - \pi C \label{eq:first-theta-C}
\end{align}
is always bosonic, i.e., that the Chern number $C$ is always even (Sec.~\ref{sec:analysis}).
The statistics of skyrmions thus never changes for the systems of interest.

Apart from a Hopf term, the low-energy effective action of $\vu{n}(x)$ includes kinetic terms such as
\begin{align}
S_{\text{kin}}[\vu{n}] &= \frac{K}{2} \int \dd[3]{x} \, \mleft(\partial_{\mu} \vu{n}\mright)^2, \label{eq:NLSM-kinetic-part}
\end{align}
where $K$ is the phase stiffness.
(The fermions are absent from the low-energy effective theory because they are gapped.)
This particular kinetic term describes antiferromagnets.
Ferromagnets have an additional spin Wess-Zumino term~\cite{Valenti1977, Jevicki1979, Klauder1979, Abanov2017}.
Since we are in 2+1D, at $T = 0$ the $K$ of this $\Ogp(3)$ non-linear sigma model may either flow under RG towards the ordered $K \to \infty$ phase or towards the disordered $K = 0$ phase, depending on the initial value of $K$~\cite{Altland2010}.
At finite temperatures, however, the Hohenberg-Mermin-Wagner theorem~\cite{Hohenberg1967, Mermin1966} forbids the spontaneous breaking of the $\SO(3)$ rotation symmetry so $K$ always flows towards zero.

In the ordered phase, the $\SU(2)$ spin rotation symmetry is broken and $\vu{n}$ fluctuates around some mean value $\vu{n}_0$.
Skyrmionic and hopfionic field configurations are strongly suppressed by their kinetic energy cost.
Hence the Hopf term is irrelevant.
That said, when the Chern number $C$ is finite, the ordered phase is a TR symmetry-breaking topological insulator~\cite{Chiu2016} that has chiral gapless fermionic edge states, in addition to the gapless spin waves of the bulk, and it experiences the anomalous Hall effect~\cite{Nagaosa2010}.
Because the Chern number vanishes for both a TR-odd $\vu{n}$ and a TR-even $\vu{n}'$ in isolation, as we show in Sec.~\ref{sec:Hopf-symmetry}, both the magnetic and the spin loop-current insulators of Fig.~\ref{fig:phase-diagram} are always topologically trivial.

The disordered phase, in contrast, is essentially an instantonic (hopfionic) gas and is thus governed by the Hopf term.
For bosonic $\theta \in 2 \pi \Z$ that are non-zero, the disordered fixed point describes an $\SU(2)$ symmetry protected topological (SPT) phase~\cite{Liu2013, Yue2023}.
This phase preserves the $\SU(2)$ spin rotation symmetry and is topologically non-trivial due to short-ranged entanglement~\cite{Gu2009, Senthil2015}.
Since $\theta = - \pi C$ for our system (Eq.~\eqref{eq:first-theta-C}), as we shall see in Sec.~\ref{sec:Hopf-gradient}, to realize this phase we shall need to somehow break TR symmetry without breaking the spin rotation symmetry.
This is done by considering the region where there is both a TR-odd magnetic $\vu{n}(x)$ and a TR-even spin loop-current $\vu{n}'(x)$ present in the system.
The TR-odd Ising variable $\vu{n}(x) \vdot \vu{n}'(x)$ may then condensed, as depicted in Fig.~\ref{fig:phase-diagram}, to give a non-zero Chern number of the band Hamiltonian (Sec.~\ref{sec:Hopf-symmetry}).
At zero temperature below this TR symmetry-breaking phase, a Hopf SPT will generically appear (Fig.~\ref{fig:phase-diagram}).
There are two fruitful ways of studying this SPT fixed point: either by looking at the boundary of the system, or by probing the system with an external $\SU(2)$ gauge field.

Due to the constructive (bosonic) interference between topological sectors, the bulk of the disordered phase continues to be gapped, as it would be in the absence of a Hopf term.
However, if a system possesses a boundary, the topological Hopf theta term of the bulk acts as a Wess-Zumino term of the edge theory.
As we shall show in Sec.~\ref{sec:Hopf-gradient}, in terms of the $\SU(2)$ group-valued field $U(x)$ that is defined through $U^{-1}(x) \, \vu{n}(x) \vdot \vb{\Pauli} \, U(x) = \Pauli^3$,
\begin{align}
Q_{\text{Hopf}} &= \frac{1}{24 \pi^2} \int \tr (U^{-1} \dd{U})^{\wedge 3},
\end{align}
where $\wedge$ is the wedge product.
On symmetry grounds, the edge theory therefore takes the form of a Wess-Zumino-Witten model of level $k = \theta / (2 \pi)$~\cite{Liu2013, Yue2023}:
\begin{align}
\begin{aligned}
S_{\text{edge}}[U] &= \int \dd[2]{x} \mleft[\frac{1}{\lambda} \tr(\partial_{\mu} U^{-1}) (\partial^{\mu} U) + \mathcal{L}_{\text{int}}\mright] \\
&\quad - \iu \frac{k}{12 \pi} \int \tr\mleft(U^{-1} \dd{U}\mright)^{\wedge 3},
\end{aligned} \label{eq:WZW}
\end{align}
where $x^0 \equiv \tau$ is imaginary time, $x^1$ goes along the spatial boundary, $\lambda$ is a coupling constant, and $\mathcal{L}_{\text{int}}$ are additional terms that break the $\SU(2)_R$ right-multiplication symmetry $U(x) \mapsto U(x) \cdot g$, where $g \in \SU(2)$.
The $\SU(2)_L$ left-multiplication symmetry $U(x) \mapsto g \cdot U(x)$ corresponds to the spin rotation symmetry and is preserved by $\mathcal{L}_{\text{int}}$.

The Wess-Zumino-Witten model (without $\mathcal{L}_{\text{int}}$) has been extensively studied~\cite{Witten1984, Knizhnik1984, Francesco1997} and it is well-established that it flows towards the gapless fixed point $\lambda = 8 \pi / \abs{k}$.
At this fixed point, the model has decoupled left-movers $J_{+}$ and right-movers $J_{-}$ whose equations of motion are
\begin{align}
\partial_{\mp} J_{\pm} &= 0,
\end{align}
where $x^{\pm} \defeq x^0 \pm \iu \, x^1$, $\partial_{\pm} \defeq \partial / \partial x^{\pm}$, and
\begin{align}
J_{+} &\defeq \begin{cases}
(\partial_{+} U) U^{-1}, & \text{for $k > 0$,} \\
- U^{-1} (\partial_{+} U), & \text{for $k < 0$,}
\end{cases} \\
J_{-} &\defeq \begin{cases}
- U^{-1} (\partial_{-} U), & \text{for $k > 0$,} \\
(\partial_{-} U) U^{-1}, & \text{for $k < 0$.}
\end{cases}
\end{align}
For both positive and negative $k$, $J_{+}$ and $J_{-}$ have different transformation properties under left-multiplication $U \mapsto g \cdot U$, one being invariant and the other being covariant, i.e., charged.
As shown in Ref.~\cite{Liu2013} (cf.\ Ref.~\cite{Xu2013}), $J_{\pm}$ will continue to be gapless even when $\SU(2)_R$ is broken by $\mathcal{L}_{\text{int}}$ because the mass term $\sim \tr U^2$ is forbidden by $\SU(2)_L$ and because the different $\SU(2)_L$ charges prevent scattering between $J_{\pm}$ (e.g., through a scattering term $\sim \tr J_{+} J_{-}$) that could also potentially gap the edge states.
The SPT Hopf phase therefore has chiral gapless edge states that carry $\SU(2)_L$ charge preferentially in one direction~\cite{Liu2013, Yue2023}.

The SPT Hopf phase can also be studied by coupling it to an external $\SU(2)$ gauge field $\mathcal{B}_{\mu}$.
This is done through the minimal substitution $U^{-1} \partial_{\mu} U \to U^{-1} \mleft(\partial_{\mu} + \iu \mathcal{B}_{\mu}\mright) U$.
For the Hopf term that describes the SPT fixed point this results in~\cite{Liu2013}:
\begin{align}
\begin{aligned}
\tr (U^{-1} \dd{U})^{\wedge 3} \to \tr\!\Big[&- \iu \mathcal{B}^{\wedge 3} - 3 \mathcal{F} \wedge \tilde{A} \\
&+ 3 \dd{(\mathcal{B} \wedge \tilde{A})} - \iu \tilde{A}^{\wedge 3}\Big],
\end{aligned}
\end{align}
where $\mathcal{F}_{\mu \nu} \defeq \partial_{\mu} \mathcal{B}_{\nu} - \partial_{\nu} \mathcal{B}_{\mu} + \iu [\mathcal{B}_{\mu}, \mathcal{B}_{\nu}]$ and $\tilde{A}_{\mu} \defeq - \iu (\partial_{\mu} U) U^{-1}$.
From this expression we see that $\mathcal{B}_{\mu}$ couples only to $\tilde{A}_{\mu}$, which corresponds to $J_{+}$ ($J_{-}$) on the boundary when $k > 0$ ($k < 0$).
After integrating out the group field $U(x)$, one finds a non-abelian $\SU(2)$ Chern-Simons term in the effective response action~\cite{Liu2013}:
\begin{align}
S_{\text{resp}}[\mathcal{B}] &= - \iu \frac{k}{4 \pi} \int \tr\mleft(\mathcal{B} \wedge \mathcal{F} - \iu \frac{1}{3} \mathcal{B} \wedge \mathcal{B} \wedge \mathcal{B}\mright).
\end{align}
There are two important implications of this.
The first is that the spin current $\mathcal{J}_{\mu}^a = \var{S_{\text{resp}}} / \var{\mathcal{B}_{\mu}^a}$ experiences the spin quantum Hall effect, with a spin-Hall coefficient that is proportional to $k = \theta / (2 \pi) = - C / 2$~\cite{Liu2013, Yue2023}.
The other is that, in the presence of a boundary, this effective response action has a gauge anomaly that can be partially integrated to only reside on the boundary.
To cancel this gauge anomaly, the edge theory therefore must have chiral gapless fermions~\cite{Liu2013, Yue2023}.
Via non-abelian bosonization~\cite{Witten1984}, these gapless fermionic edge states correspond to the same gapless edge states described by the Wess-Zumino-Witten model of Eq.~\eqref{eq:WZW}.

\section{Spin-ordered insulators} \label{sec:analysis}
In this part, we analyze the situation depicted in Fig.~\ref{fig:phase-diagram}: that of an intersection between a magnetic and spin loop-current insulator.
As we shall demonstrate, a Hopf term generically appears here, resulting in a Hopf SPT at zero temperature.
This Hopf SPT, which we analyzed in Sec.~\ref{sec:Hopf-properties}, is characterized by chiral edge states and a quantized spin-Hall conductance.

Although radiatively inducing a Hopf term through coupling to fermionic spins has been studied in many contexts~\cite{Volovik1989, Moon1995, Iordanskii1997, Apel1997, Ray1999, Sengupta2000, Yakovenko1990, Yakovenko1991, Grover2008, Yue2023, Yue2025}, the concrete setting and approach that we are employing here has not been done.
Moreover, given recent advances in engineering magnetic heterostructures~\cite{Tokura2019, Gibertini2019, Huang2020, Blei2021, Petrovic2025}, there is hope for realizing our proposal.
To guide experimentalists, we carry out a detailed symmetry analysis (Sec.~\ref{sec:Hopf-symmetry}) of a very general spin-ordered insulator model (Sec.~\ref{sec:general-model}) to see what are the necessary conditions for the appearance of a radiatively-induced Hopf term (Sec.~\ref{sec:Hopf-gradient}).
We find that simultaneous magnetic and spin loop-current order is needed, in addition to the absence of reflection symmetries.
In the last Sec.~\ref{sec:lattice-model}, we provide lattice model realizations of our proposal.

\subsection{Spin-ordered insulator model} \label{sec:general-model}
Even in purely fermionic systems, bosonic 3-component fields $\vb{\phi}(x)$ can emerge as the effective degrees of freedom at low energies~\cite{Hertz1976, Millis1993, Moriya1985, Auerbach1994, Abanov2003, Altland2010}.
This happens naturally when a Fermi liquid condenses into a spin-ordered insulating phase with a pseudovector order parameter $\vb{\phi}(x)$.
Since the fermions are gapped, only the fluctuations of $\vb{\phi} = \bar{\phi} \, \vu{n}$ are relevant at low energies.
In the absence of spin-orbit coupling, the spin rotation symmetry implies that long-wavelength fluctuations of the orientation unit-vector field $\vu{n}(x)$ are energetically inexpensive.
Physically, this phase is characterized by magnetism (spin loop currents) for TR-odd (TR-even) $\vb{\phi}(x)$.

The order parameter $\vb{\phi}(x)$ is formally introduced by performing a Hubbard-Stratonovich transformation~\cite{Altland2010} on the dominant four-fermion interaction channel.
By assumption, in our system the dominant interaction is
\begin{align}
S_{\text{int}}[\psi] &= - \frac{1}{2} \int_{0}^{\beta} \dd{\tau} \sum_{\vb{R}_1 \vb{R}_2} V(\vb{R}_1 - \vb{R}_2) \, \vb{S}(\vb{R}_1) \vdot \vb{S}(\vb{R}_2), \label{eq:f-interaction}
\end{align}
where $\vb{R}$ go over the real-space lattice,
\begin{align}
S^a(\vb{R}) &\defeq \frac{1}{2} \sum_{\vb{\delta}} \psi^{\dag}(\vb{R} + \vb{\delta}) \Gamma(\vb{\delta}) \Pauli^a \psi(\vb{R}) + \Hc
\end{align}
is a spin-like Hermitian operator, $\vb{S}^{\dag}(\vb{R}) = \vb{S}(\vb{R})$, and $V(\vb{R}) = V^{*}(\vb{R}) = V(-\vb{R})$ is the interaction.
Here $\vb{\delta}$ go over lattice neighbors, $\psi(\vb{R})$ is the column-vector of fermionic field (annihilation) operators, $\Gamma(\vb{\delta})$ are orbital matrices, and $\Pauli^a$ are Pauli spin matrices.
After a Hubbard-Stratonovich transformation:
\begin{gather}
\begin{gathered}
S_{\text{int}}[\psi, \vb{\phi}] = \int_{0}^{\beta} \dd{\tau} \sum_{\vb{R}} \vb{S}(\vb{R}) \vdot \vb{\phi}(\vb{R}) \qquad \\
\quad + \frac{1}{2} \int_{0}^{\beta} \dd{\tau} \sum_{\vb{R}_1 \vb{R}_2} V^{-1}(\vb{R}_1 - \vb{R}_2) \, \vb{\phi}(\vb{R}_1) \vdot \vb{\phi}(\vb{R}_2),
\end{gathered}
\end{gather}
where $\vb{\phi}$ is a real 3-component field and $V^{-1}$ is the matrix inverse, $\sum_{\vb{R}_2} V(\vb{R}_1 - \vb{R}_2) V^{-1}(\vb{R}_2 - \vb{R}_3) = \Kd_{\vb{R}_1 \vb{R}_3}$.

When $\vb{S}(\vb{R}) \vdot \vb{\phi}$ gaps the fermions, pushing the occupied states to lower energies, it may become energetically favorable for $\vb{\phi} = \bar{\phi} \, \vu{n}$ to condense.
For our system, we shall always assume that it is in a regime where condensation takes place.
For reasons that shall become apparent in Sec.~\ref{sec:Hopf-symmetry}, we shall also assume that there is another $\vb{\phi}' = \bar{\phi}' \, \vu{n}'$ which is TR even (spin loop-current), in contrast to the TR odd (magnetic) $\vb{\phi} = \bar{\phi} \, \vu{n}$ (Fig.~\ref{fig:phase-diagram}).
The Hubbard-Stratonovich transformation for $\vb{\phi}'$ proceeds analogously to $\vb{\phi}$.
The fermionic action now acquires the form:
\begin{align}
\begin{aligned}
S_{\text{ferm}}[\psi, \vu{n}] = \int_{0}^{\beta} \dd{\tau} \sum_{\vb{R}_1 \vb{R}_2} \bar{\psi}(\vb{R}_1) \Big[\partial_{\tau} + H(\vb{R}_1-\vb{R}_2) \\
+ \, \Gamma(\vb{R}_1-\vb{R}_2) \tfrac{1}{2} \big(\vu{n}(\vb{R}_1) + \vu{n}(\vb{R}_2)\big) \vdot \vb{\Pauli} \\
+ \, \Gamma'(\vb{R}_1-\vb{R}_2) \tfrac{1}{2} \big(\vu{n}'(\vb{R}_1) + \vu{n}'(\vb{R}_2)\big) \vdot \vb{\Pauli}\Big] \psi(\vb{R}_2),
\end{aligned} \label{eq:gen-f-action}
\end{align}
where $H(\vb{\delta}) = H^{\dag}(-\vb{\delta})$ is the band Hamiltonian.
Here $\Gamma(\vb{\delta}) = \Gamma^{\dag}(-\vb{\delta})$ and $\Gamma'(\vb{\delta}) = \Gamma^{\prime \dag}(-\vb{\delta})$ have absorbed the amplitudes $\bar{\phi}$ and $\bar{\phi}'$ of the order parameters $\vb{\phi} = \bar{\phi} \, \vu{n}$ and $\vb{\phi}' = \bar{\phi}' \, \vu{n}'$, respectively.
Since the Hamiltonian $H(\vb{\delta})$ describes the already renormalized Fermi liquid quasi-particles, we shall not include further fermionic interactions.
To enable gapless fluctuations of $\vu{n}$ and $\vu{n}'$, we shall assume no spin-orbit coupling, $[H, \Pauli^a] = 0$.

\subsection{Radiatively-induced Hopf terms} \label{sec:Hopf-gradient}
The fermions described by Eq.~\eqref{eq:gen-f-action} may induce various terms in the effective action of $\vu{n}$ and $\vu{n}'$, including Hopf terms.
For the moment let us focus on $\vu{n}$ and omit $\vu{n}'$; we shall reintroduce it later.
In principle, all these terms are contained in the exact expression
\begin{align}
S_0[\vu{n}] &= - \Tr \log\mleft(\partial_{\tau} + H + \Gamma \, \vu{n} \vdot \vb{\Pauli}\mright)
\end{align}
that one obtains by integrating out the fermions.
However, to get something manageable out of this expression, further manipulations and approximations are needed, namely, an expansion in the gradients of $\vu{n}$.
Hopf terms that are generated and that can be captured by a gradient expansion we shall call \emph{radiatively induced}~\cite{Dunne1999}.

The gradient expansion is performed by rotating the fermionic spinors along $\vu{n}(x)$ with the help of an $\SU(2)$ group-valued field $U(x)$ that is defined through~\cite{Volovik1989}:
\begin{align}
U^{-1}(x) \, \vu{n}(x) \vdot \vb{\Pauli} \, U(x) &= \Pauli^3. \label{eq:SU2-param-def}
\end{align}
There is a $U(x) \mapsto U(x) \exp(\iu \chi(x) \Pauli^3)$ $\Ugp(1)$ gauge freedom inherit to this definition.
Gradients of $\vu{n}$ now correspond to the non-abelian $\SU(2)$ vector potential
\begin{align}
A_{\mu} &\equiv \sum_{a=1}^{3} A_{\mu}^a \Pauli^a \equiv \vb{A}_{\mu} \vdot \vb{\Pauli} \defeq - \iu \, U^{-1} (\partial_{\mu} U),
\end{align}
which is gauge-free in the sense that it has a vanishing non-abelian $\SU(2)$ gauge field strength or curvature $F_{\mu\nu}$, as follows from $U^{-1} [\partial_{\mu}, \partial_{\nu}] U = 0$:
\begin{align}
\begin{aligned}
F_{\mu\nu} \equiv \vb{F}_{\mu\nu} \vdot \vb{\Pauli} \defeq \partial_{\mu} A_{\nu} - \partial_{\nu} A_{\mu} + \iu \mleft[A_{\mu}, A_{\nu}\mright] &= 0, \\
\vb{F}_{\mu\nu} = \partial_{\mu} \vb{A}_{\nu} - \partial_{\nu} \vb{A}_{\mu} - 2 \vb{A}_{\mu} \vcross \vb{A}_{\nu} &= \vb{0}.
\end{aligned} \label{eq:gauge-free}
\end{align}
$A_{\mu}$ is related to the vector potential $a_{\mu}$ of Eq.~\eqref{eq:vec-pot-a} via
\begin{align}
a_{\mu} &= \frac{1}{2 \pi} A_{\mu}^{3}, \label{eq:a-A3-rel}
\end{align}
from which we see that one cannot introduce a global $U(x)$ whenever skyrmions are present (Sec.~\ref{sec:theta-quant}).
This poses no problem for the gradient expansion because it is performed locally around a point.
The Hopf invariant~\eqref{eq:Hopf-CS} in terms of $U$ is simply the winding number of mappings from the spacetime 3-sphere $\mathcal{S}^3$ to $\SU(2) \cong \mathcal{S}^3$:
\begin{align}
\begin{aligned}
Q_{\text{Hopf}} &= \frac{1}{24 \pi^2} \int \tr (U^{-1} \dd{U})^{\wedge 3} \\
&= \frac{1}{12 \pi^2} \int \dd[3]{x} \, \LCs^{\mu \nu \rho} \vb{A}_{\mu} \vdot (\vb{A}_{\nu} \vcross \vb{A}_{\rho}).
\end{aligned} \label{eq:Hopf-U}
\end{align}
Proof and further properties of $U$ are provided in App.~\ref{sec:U-properties}.

When $\vu{n}(x)$ slowly varies in space and time, we are justified in expanding the action~\eqref{eq:gen-f-action} in gradients using $\psi(x) \mapsto U(x) \psi(x)$.
The result is (App.~\ref{sec:gradient-formula})
\begin{align}
\begin{aligned}
S_{\text{ferm}}[\psi, U] = \sum_{k, q} \bar{\psi}_{k+q/2} \Big[&G_k^{-1} \Kd_q + \mathcal{V}_{k,q}^{(1)} \\[-10pt]
&+ \mathcal{V}_{k,q}^{(2)} + \mathcal{V}_{k,q}^{(3)} + \cdots\Big] \psi_{k-q/2},
\end{aligned}
\end{align}
where $x \equiv (\tau, \vb{R})$, $k \equiv k^{\mu} \equiv k_{\mu} = (- \omega_k, \vb{k})$, $q = (- \omega_q, \vb{q})$, the frequencies are Matsubara, the momenta go over the first Brillouin zone, and
\begin{align}
G_k^{-1} &= \iu k_0 + H_{\vb{k}} + \Gamma_{\vb{k}} \Pauli^3, \\
\mathcal{V}_{k,q}^{(1)} &= \frac{1}{2} \mleft\{\pdv{G^{-1}_k}{k_{\mu}}, \ev{A_{\mu}}_q\mright\}, \\
\mathcal{V}_{k,q}^{(2)} &= \frac{1}{4} \mleft\{\pdv{^2 G^{-1}_k}{k_i \partial k_j}, \ev{A_i A_j}_q\mright\}, \\
& \hspace{-19pt} \begin{aligned}
\mathcal{V}_{k,q}^{(3)} &= \frac{1}{12} \bigg\{\pdv{^3 G^{-1}_k}{k_i \partial k_j \partial k_{\ell}}, \ev{A_i A_j A_{\ell}}_q \\
&\hspace{-18pt} + \ev{\tfrac{1}{2} \iu (\partial_i A_j) A_{\ell} - \tfrac{1}{2} \iu A_i (\partial_j A_{\ell}) - \tfrac{1}{4} \partial_i \partial_j A_{\ell}}_q\bigg\}.
\end{aligned}
\end{align}
Here the curly brackets are anticommutators and summations over $\mu \in \{0, 1, 2\}$ and $i, j, \ell \in \{1, 2\}$ are implicit.
Since the fermions are defined on a lattice, the path integral measure is invariant under $\psi(x) \mapsto U(x) \psi(x)$, i.e., there is no anomaly.

To find the Hopf term, we need to identify which terms include $\LCs^{\mu \nu \rho} \vb{A}_{\mu} \vdot (\vb{A}_{\nu} \vcross \vb{A}_{\rho})$, as in Eq.~\eqref{eq:Hopf-U}, or $\LCs^{\mu \nu \rho} \vb{A}_{\mu} \vdot \partial_{\nu} \vb{A}_{\rho}$, due to Eq.~\eqref{eq:gauge-free}, in the expansion of
\begin{align}
S_0[U] &= - \Tr \log G^{-1} - \Tr \log\mleft(\one + G \mathcal{V}^{(1)} + \cdots\mright). \label{eq:SU-Tr-log}
\end{align}
Here one should keep in mind that derivatives of $\vb{A}_{\mu}$ can not only arise explicitly through $\mathcal{V}^{(n)}$, but also indirectly when in the expansion in slow momenta we find terms proportional to the slow momenta.

The potentially relevant terms are
\begin{align}
\begin{aligned}
S_1 = &- \tfrac{1}{3} \Tr G \mathcal{V}^{(1)} G \mathcal{V}^{(1)} G \mathcal{V}^{(1)} \\
&+ \Tr G \mathcal{V}^{(1)} G \mathcal{V}^{(2)} - \Tr G \mathcal{V}^{(3)}
\end{aligned}
\end{align}
to zeroth order in slow momenta and
\begin{align}
S_2 = \tfrac{1}{2} \Tr G \mathcal{V}^{(1)} G \mathcal{V}^{(1)} - \Tr G \mathcal{V}^{(2)}
\end{align}
to first order in slow momenta.
In $S_1$, the $\Tr G \mathcal{V}^{(3)}$ term includes only spatial components and derivatives and therefore cannot contribute to the Hopf term.
In $\mathcal{V}^{(2)}$, the fact that $A_i A_j$ contracts with the symmetric tensor $\partial_i \partial_j G_k = \partial_j \partial_i G_k$ means that we effectively have $\tfrac{1}{2} \{A_i, A_j\} = \vb{A}_i \vdot \vb{A}_j \, \one$.
Hence the $\Tr G \mathcal{V}^{(1)} G \mathcal{V}^{(2)}$ term in $S_1$ has the wrong index structure for the Hopf term.
The $\Tr G \mathcal{V}^{(2)} = \sum_k \tr G_k \mathcal{V}^{(2)}_{k,q=0}$ term in $S_2$ has no slow momentum ($q = 0$) so it drops out as well.
The relevant terms are therefore:
\begin{align}
&\begin{aligned}
S_1' = - \frac{1}{3} &\sum_{k, q_1, q_2} \tr\!\Big(G_k \mathcal{V}^{(1)}_{k+q_1/2+q_2/2, -q_1-q_2} \\
&G_{k+q_1+q_2} \mathcal{V}^{(1)}_{k+q_1+q_2/2, q_2} G_{k+q_1} \mathcal{V}^{(1)}_{k+q_1/2, q_1}\Big),
\end{aligned} \\
&S_2' = \frac{1}{2} \sum_{k,q} \tr G_k \mathcal{V}^{(1)}_{k+q/2, -q} G_{k+q} \mathcal{V}^{(1)}_{k+q/2, q},
\end{align}
which are the same ones that have been originally identified for a simpler continuum model~\cite{Volovik1989}.
After expanding in the slow momenta $q$, $q_1$, and $q_2$, we thus find
\begin{align}
S_1'' &= \frac{- 1}{3 L^2 \beta} \sum_k \mleft(\tr X_{\mu}^a X_{\nu}^b X_{\rho}^c\mright)_k \int \dd[3]{x} \mleft(A_{\mu}^a A_{\nu}^b A_{\rho}^c\mright)_x, \\
S_2'' &= \frac{\iu}{2 L^2 \beta} \sum_k \mleft(\tr X_{\mu}^a X_{\rho} X_{\nu}^b\mright)_k \int \dd[3]{x} \mleft(A_{\mu}^a \partial_{\rho} A_{\nu}^b\mright)_x, \label{eq:S2''}
\end{align}
where $L^2$ is the total area of the system and
\begin{align}
X_{\mu}(k) \defeq G_k \pdv{G_k^{-1}}{k_{\mu}}, \hspace{7pt}
X_{\mu}^a(k) \defeq G_k \frac{1}{2} \mleft\{\pdv{G_k^{-1}}{k_{\mu}}, \Pauli^a\mright\}. \label{eq:Xmua-def}
\end{align}

Equating $S_1'' + S_2'' = - \iu \theta Q_{\text{Hopf}} + \cdots$ yields the final expression for the Hopf $\theta$ angle:
\begin{align}
\theta &= - \frac{1}{24 \pi} \int \dd[3]{k} \, \LCs^{\mu \nu \rho} \tr\mleft[\frac{\iu}{3} \LCs_{abc} X_{\mu}^a X_{\nu}^b X_{\rho}^c + X_{\mu} X_{\nu}^a X_{\rho}^a\mright]. \label{eq:final-theta}
\end{align}
Here the $T = 0$ thermodynamic limit has been taken, the $k_0 = - \omega$ integral goes from $- \infty$ to $\infty$, $\vb{k} = (k_1, k_2)$ goes over the first Brillouin zone, and summations over $\mu, \nu, \rho \in \{0, 1, 2\}$ and $a, b, c \in \{1, 2, 3\}$ are implicit.
This is the main result of the current section.

This formula for the $\theta$ angle, in fact, applies more generally to topological theta terms in 2+1D that are associated with winding numbers of $\SU(2)$ group-valued fields (Eq.~\eqref{eq:Hopf-U}).
Let us consider a system that simultaneously orders in two spin channels whose order parameters have orientations $\vu{n}$ and $\vu{n}'$, as originally in Eq.~\eqref{eq:gen-f-action}.
At low energies, the fluctuations in the relative orientation $\vu{n} \vdot \vu{n}' = \cos \alpha$ will becomes massive, just like amplitude fluctuations.
An $\SU(2)$ group-valued field defined through
\begin{align}
\begin{aligned}
U^{-1}(x) \, \vu{n}(x) \vdot \vb{\Pauli} \, U(x) &= \Pauli^3, \\
U^{-1}(x) \, \vu{n}'(x) \vdot \vb{\Pauli} \, U(x) &= \cos \alpha \, \Pauli^3 + \sin \alpha \, \Pauli^1
\end{aligned}
\end{align}
will therefore describe the low-energy physics, assuming no spin-orbit coupling as before.
Notice how there is no residual $\Ugp(1)$ gauge freedom left in this definition of $U(x)$, unless $\alpha = 0$ or $\pi$.
The $\theta$ resulting from the fermionic action~\eqref{eq:gen-f-action} will now be given by Eq.~\eqref{eq:final-theta}, but with a modified
\begin{align}
G_k^{-1} &= \iu k_0 + H_{\vb{k}} + \Gamma_{\vb{k}} \Pauli^3 + \Gamma_{\vb{k}}' \mleft(\cos \alpha \, \Pauli^3 + \sin \alpha \, \Pauli^1\mright). \label{eq:modified-Gk}
\end{align}

When $\vu{n}$ and $\vu{n}'$ have opposite TR signs, relative angles of $\alpha$ and $\pi - \alpha$ have the same energy and $\vu{n} \vdot \vu{n}' = \cos \alpha$ is a TR-odd Ising variable that may condense at finite temperatures (Fig.~\ref{fig:phase-diagram}).
Within the TR symmetry-breaking phase, we cannot make rigorous general statements regarding $\alpha$, except that $\alpha \neq \pi/2$.
However, we can give two heuristic arguments for why $\alpha$ should generically equal $0$ or $\pi$, as needed for a Hopf term proper.
On the one hand, when the gapping of the Fermi liquid is small, only retaining the lowest-order term in the free energy expansion $F = F_0 + a \mleft(\vu{n} \vdot \vu{n}'\mright)^2 + \cdots$, with $a < 0$, gives minima at $\alpha = 0, \pi$.
On the other hand, due to the different TR signs, the nodes of the momentum-dependent gapping are located at different points for $\vu{n}$ and $\vu{n}'$, making it energetically preferable to have (anti-)parallel $\vu{n}$ and $\vu{n}'$ so as to maximize the gapping.
This latter argument we have verified on the lattice models of Sec.~\ref{sec:lattice-model}.

For the case of a Hopf term proper, the $\Ugp(1)$ gauge freedom $U \mapsto U \Elr^{\iu \chi \Pauli^3}$ implies that $[G_k^{-1}, \Pauli^3] = 0$.
By rewriting Eq.~\eqref{eq:final-theta} into
\begin{align}
\theta &= - \frac{1}{24 \pi} \int \dd[3]{k} \, \LCs^{\mu \nu \rho} \tr\mleft(\iu X_{\mu}^3 \{X_{\nu}^1, X_{\rho}^2\} + X_{\mu} X_{\nu}^a X_{\rho}^a\mright),
\end{align}
it is easy to see that the first term cancels with the $a = 1, 2$ terms of the second term, leaving only $X_{\mu} X_{\nu}^3 X_{\rho}^3 = X_{\mu} X_{\nu} X_{\rho}$.
Hence~\cite{Volovik1989}:
\begin{align}
\theta &= - \pi C, \label{eq:theta-Chern}
\end{align}
where $C \in \Z$ is the Chern number
\begin{align}
\begin{aligned}
C &= \frac{1}{24 \pi^2} \int \dd[3]{k} \, \LCs^{\mu \nu \rho} \tr X_{\mu} X_{\nu} X_{\rho} \\
&= \frac{1}{8 \pi^2} \int \dd[3]{k} \tr X_0 \mleft[X_1, X_2\mright].
\end{aligned} \label{eq:Chern}
\end{align}
In agreement with the previously provided proof of Sec.~\ref{sec:theta-quant}, we find that $\theta$ is quantized.

If we diagonalize $\mathscr{H}_{\vb{k}} = G_k^{-1} - \iu k_0$, $\mathscr{H}_{\vb{k}} \ket{n} = E_n \ket{n}$, and carry out the frequency integral in Eq.~\eqref{eq:Chern}, we obtain the traditional expression for the Chern number:
\begin{align}
C &= \frac{\iu}{2 \pi} \int \dd[2]{k} \sum_n \Theta(-E_n) \, \Omega_n.
\end{align}
Here $\Theta$ is the Heaviside function ($T = 0$) and $\Omega_n$ is the Berry curvature which appears here through the Mead-Truhlar formula ($\partial_i \defeq \partial / \partial k_i$):
\begin{align}
\Omega_n &\defeq \sum_{m \neq n} \frac{\LCs^{0ij} \mel{n}{\partial_i \mathscr{H}_{\vb{k}}}{m} \mel{m}{\partial_j \mathscr{H}_{\vb{k}}}{n}}{(E_n - E_m)^2}.
\end{align}

If we were at finite $T$, in Eq.~\eqref{eq:Chern} instead of a frequency integral we would have a sum over fermionic Matsubara frequencies.
One would then readily find that $C$ is not an integer anymore, seemingly violating invariance under large gauge transformations (Sec.~\ref{sec:theta-quant}).
This is, in fact, a well-known problem with finite-$T$ gradient expansions~\cite{Sengupta2000}.
As observed in Ref.~\cite{Sengupta2000}, the polarization bubble that one expands around $q = 0$ to get the $S_2''$ of Eq.~\eqref{eq:S2''} is non-analytic at finite temperatures, i.e., it is sensitive to the order of the $q_0 \to 0$ and $\vb{q} \to \vb{0}$ limits.
A comparison with radiatively-induced Chern-Simons terms in 2+1D QED is informative~\cite{Dunne1999, Salcedo2002, Deser1997, Fosco1997, *Fosco1997-E}.
There, it is known that the source of the error lies in expanding in the temporal vector potential component $A_0$~\cite{Dunne1999, Salcedo2002}.
The Ward identity associated with large gauge transformations mixes orders in $A_0$~\cite{Deser1997} and terms that are non-local in time are needed to restore large gauge invariance~\cite{Dunne1999, Salcedo2002, Deser1997, Fosco1997, *Fosco1997-E}.
The same is expected for finite-$T$ Hopf terms, although we shall not pursue this topic further here.
Let us just note that, unlike for 2+1D QED, in our system one cannot choose a gauge in which $A_0$ is constant so the gradient expansion cannot be modified, as has elegantly been done in Ref.~\cite{Salcedo2002} for QED, to correctly describe finite temperatures.

\subsection{Symmetry considerations} \label{sec:Hopf-symmetry}
Although formula~\eqref{eq:theta-Chern} for the Hopf $\theta$ has been written down and applied to a variety of systems~\cite{Volovik1989, Moon1995, Iordanskii1997, Apel1997, Ray1999, Sengupta2000, Yakovenko1990, Yakovenko1991, Hlousek1990}, the necessary conditions for $\theta$ as given by Eq.~\eqref{eq:theta-Chern}, that is Eq.~\eqref{eq:final-theta}, to be finite have not yet been spelled out in detail.
As we shall see, there are subtle differences in the symmetry properties of formula~\eqref{eq:theta-Chern}, as oppose to those of the Hopf phase factor $\Elr^{\iu \theta Q_{\text{Hopf}}}$, which indicate limitations of the gradient expansion.

Let us start by considering the symmetry properties of the Hopf invariant~\eqref{eq:Hopf-CS}.
The corrections of Eq.~\eqref{eq:Hopf-CS-v2} have the same symmetry properties as Eq.~\eqref{eq:Hopf-CS}, as one may show.
Since $f^{\mu} \sim \LCs^{\mu \nu \rho} \vu{n} \vdot \mleft(\partial_{\nu} \vu{n} \vcross \partial_{\rho} \vu{n}\mright)$, $\vb{f}$ is a pseudoscalar in $\vu{n}$ space and a pseudovector ($\sim \grad \vcross \grad$) in spacetime indices.
From $\vb{f} = \grad \vcross \vb{a}$, we see that $\vb{a}$ is also a pseudoscalar in $\vu{n}$ space, but a spacetime vector.
Hence $Q_{\text{Hopf}} = \int \dd[3]{x} \, \vb{a} \vdot \vb{f}$ is invariant under all $\vu{n} \mapsto R \vu{n}$ transformations, where $R \in \Ogp(3)$, and transforms like a spacetime pseudoscalar under spacetime transformations that do not act on $\vu{n}$.
In particular, $Q_{\text{Hopf}}$ is odd under spacetime inversion $x \mapsto - x$, time reversal $(\tau, \vb{r}) \mapsto (-\tau, \vb{r})$, and space reflections $(\tau, \vb{r}) \mapsto (\tau, \vb{r} - 2 \vb{r} \vdot \vu{v} \, \vu{v})$ across all in-plane axes $\vu{v}$.
$Q_{\text{Hopf}}$ is invariant under all spacetime rotations, including parity $(\tau, \vb{r}) \mapsto (\tau, -\vb{r})$ that in 2+1D is identical to a $180^{\circ}$ in-plane rotation.

Point group operations and time reversal (TR) act simultaneously on spacetime and $\vu{n}$.
Since $Q_{\text{Hopf}}$ is invariant under all $\vu{n} \mapsto R \vu{n}$, the just-stated transformation rules for spacetime transformations carry over to actual point group and TR symmetry operations.
All in all, $Q_{\text{Hopf}}$ is even under parity $P$ and in-plane rotations $C_{\varphi}$ by an angle $\varphi$, but odd under TR $\Theta$ and in-plane reflections $\Sigma_{\vu{v}}$ across an axis $\vu{v}$:
\begin{align}
\begin{aligned}
P, C_{\varphi} &\colon \quad Q_{\text{Hopf}} \mapsto + Q_{\text{Hopf}}, \\
\Theta, \Sigma_{\vu{v}} &\colon \quad Q_{\text{Hopf}} \mapsto -Q_{\text{Hopf}}.
\end{aligned}
\end{align}
For layered systems embedded in 3D space, one finds that $Q_{\text{Hopf}}$ is even under out-of-plane reflections since they only act on $\vu{n}$, but odd under $180^{\circ}$ rotations around in-plane axes since they are, up to parity, the same as in-plane reflections.

If a system possesses a symmetry under which $Q_{\text{Hopf}}$ is odd, does this mean that there cannot be a Hopf term~\eqref{eq:Hopf-th-term}, i.e., that $\theta = 0$?
Not necessarily, because it is the topological phase factor $\Elr^{- S_{\text{Hopf}}} = \Elr^{\iu \theta Q_{\text{Hopf}}}$ that needs to be invariant, and not the topological action itself.
Given that $Q_{\text{Hopf}} \in \Z$ for bulk systems, the symmetry condition
\begin{align}
\Elr^{\iu \theta Q_{\text{Hopf}}} &= \Elr^{- \iu \theta Q_{\text{Hopf}}} \label{eq:theta-sym-cond}
\end{align}
is satisfied not only by $\theta = 0$, but also by any $\theta$ that is a multiple of $\pi$.
Hence Hopf terms can arise as long as they are quantized.

Now let us consider when is $\theta$ as given by Eq.~\eqref{eq:final-theta} or~\eqref{eq:theta-Chern} finite.
Call the integrand $w(k) = w(k_0, \vb{k})$ so that:
\begin{align}
\theta &= \int \dd[3]{k} \, w(k).
\end{align}
The properties of $w(k)$ turn out to be the same for both expressions~\eqref{eq:final-theta} and~\eqref{eq:theta-Chern} so we shall not differentiate between the two.
Consistency between parity and $180^{\circ}$ rotations requires that $\vu{n}$ transforms like a pseudovector, which we henceforth assume.
As for TR, we shall consider both even ($p_{\Theta} = +1$, spin loop-current) and odd ($p_{\Theta} = -1$, magnetic) $\vu{n}$.

That the Hamiltonian $\mathscr{H}_{\vb{k}} = G_k^{-1} - \iu k_0$ is Hermitian implies that:
\begin{align}
\mleft[w(k_0, \vb{k})\mright]^{*} &= w(-k_0, \vb{k}).
\end{align}
Assuming that $\mathscr{H}_{\vb{k}} = H_{\vb{k}} + \Gamma_{\vb{k}} \Pauli^3 + \Gamma_{\vb{k}}' \mleft(\cos \alpha \, \Pauli^3 + \sin \alpha \, \Pauli^1\mright)$ as in Eq.~\eqref{eq:modified-Gk}, that $\vu{n}$ and $\vu{n}'$ have the same TR sign, and that TR acts as $\Theta^{-1} \psi_{\vb{k}} \Theta = \iu \Pauli^2 \psi_{-\vb{k}}$, one finds that:
\begin{align}
\mleft[G_k^{-1}\mright]^{*} &= M_{\Theta}^{-1} G_{-k}^{-1} M_{\Theta},
\end{align}
where $M_{\Theta} = \iu \Pauli^2$ for $p_{\Theta} = +1$, and $M_{\Theta} = \one$ for $p_{\Theta} = -1$.
Either way, TR implies $\mleft[w(k)\mright]^{*} = - w(-k)$ and
\begin{align}
w(k_0, -\vb{k}) &= - w(k_0, \vb{k}).
\end{align}
Thus $\theta = 0$ in the presence of TR symmetry, unless $\vu{n}$ and $\vu{n}'$ have opposite TR signs.
This explains our interest in the crossing between spin loop-current and magnetic order (Fig.~\ref{fig:phase-diagram}).

Next, we analyze point group symmetries.
Assume the most general symmetry transformation rules:
\begin{align}
g^{-1} \psi_{\vb{k}} g &= M_{\vb{k}}(g) D(g) \psi_{R^{-1}(g) \vb{k}},
\end{align}
where $g \in \{P, C_{\varphi}, \Sigma_{\vu{v}}\}$, $M_{\vb{k}}(g)$ are unitary orbital matrices, $D(g)$ are Wigner spin-$\tfrac{1}{2}$ matrices, and
\begin{align}
\begin{aligned}
R(P) &= - \one, \\
R(C_{\varphi}) &= \begin{pmatrix} \cos \varphi & - \sin \varphi \\ \sin \varphi & \cos \varphi \end{pmatrix}, \\
R(\Sigma_{\vu{v}}) &= \one - 2 \vu{v} \vu{v}^{\intercal}
\end{aligned}
\end{align}
are in-plane vector transformation matrices.
In the presence of $g$ symmetry:
\begin{align}
M_{\vb{k}}^{-1}(g) G_{k_0, \vb{k}}^{-1} M_{\vb{k}}(g) &= G_{k_0, R^{-1}(g) \vb{k}}^{-1}.
\end{align}
After some algebra, one finds that
\begin{align}
w(k_0, R(g) \vb{k}) &= \det R(g) \, w(k_0, \vb{k}) + \partial_i K_i(k_0, \vb{k}),
\end{align}
where $\partial_i \defeq \partial / \partial k_i$, $i \in \{1, 2\}$, and $K_i(k)$ is a complicated boundary term that depends on momentum derivatives of $M_{\vb{k}}(g)$.
For momentum-independent $M_{\vb{k}}(g) = M(g)$, $K_i = 0$.
Hence reflections $\Sigma_{\vu{v}}$ imply that $\theta = 0$, whereas parity $P$ and rotations $C_{\varphi}$ are consistent with a finite Hopf $\theta$.
That the Chern number vanishes in the presence of reflection symmetry agrees with the classification of crystalline topological insulators; see Class~A in Table~VIII of Ref.~\cite{Chiu2016}.

In summary, we have found that $\theta$ as given by formulas~\eqref{eq:final-theta} and~\eqref{eq:theta-Chern} vanishes when there is TR $\Theta$ or reflection $\Sigma_{\vu{v}}$ symmetry in the system.
As a concrete example, the Dirac Hamiltonian $\mathscr{H}_{\vb{k}} = k_i \uptau^i + m \uptau^3 \Pauli^3$, where $\uptau^a$ are orbital Pauli matrices, possess both TR symmetry with $M_{\Theta} = (\iu \uptau^2) (\iu \Pauli^2)$ and reflection symmetries with $M(\Sigma_{\vu{v}}) = (\hat{v}_i \uptau^i) (\hat{v}_j \Pauli^j)$.
Consequently, when Eq.~\eqref{eq:theta-Chern} is applied, one finds that $\theta = 0$, as expected.
However, the general argument based on Eq.~\eqref{eq:theta-sym-cond} told us that $\theta$ merely has to be quantized in the presence of $\Theta$ and $\Sigma_{\vu{v}}$, and not necessarily vanish.
Moreover, non-perturbative arguments find that $\theta = \pi$ for this Dirac Hamiltonian~\cite{Abanov2000, Abanov2001, Abanov2000-p2}.
So there are strong indications that gradient expansions, at least when carried out using the standard methods of Sec.~\ref{sec:Hopf-gradient}, cannot capture Hopf terms in the presence of $\Theta$ and $\Sigma_{\vu{v}}$ symmetry.

How should we understand this limitation of gradient expansions?
In the case of Dirac fermions, the change of path integral variables $\psi(x) \mapsto U(x) \psi(x)$ might be troublesome if we have an anomaly.
Since $U(x)$ acts only in (iso)spin space, it commutes with $\gamma^0$ and it follows that there is no anomaly.
For the lattice fermions we considered, there definitely is no anomaly.
A related worry is that $\psi'(x) = U(x) \psi(x)$ might not satisfy the correct anti-periodic boundary conditions $\psi'(\beta, \vb{r}) = - \psi'(0, \vb{r})$, but this is easily fixed with a gauge transformation $U(x) \mapsto U(x) \Elr^{\iu \chi(x) \Pauli^3}$.
The most likely explanation for why the gradient expansion of Sec.~\ref{sec:Hopf-gradient} cannot capture all Hopf terms is that some phase factors are lost when the $\Tr \log$ of Eq.~\eqref{eq:SU-Tr-log} is perturbatively expanded.
This explanation is also consistent with Eq.~\eqref{eq:theta-sym-cond}: if the condition $\Elr^{\iu \theta Q_{\text{Hopf}}} = \Elr^{- \iu \theta Q_{\text{Hopf}}}$ effectively reduces to $\iu \theta Q_{\text{Hopf}} = - \iu \theta Q_{\text{Hopf}}$ because of the perturbative nature of the gradient expansion, then we shall of course find that $\theta = 0$ whenever there is $\Theta$ or $\Sigma_{\vu{v}}$ symmetry.

The simple form of Dirac Hamiltonians enables one to calculate the Hopf term, either perturbatively by embedding $\vu{n}$ in a larger manifold~\cite{Abanov2000, Abanov2001} or by extracting $\theta$ from a specific calculable field configuration~\cite{Abanov2000-p2}.
The manipulations of Ref.~\cite{Abanov2000} cannot be adapted to general dispersions, however, and other non-perturbative methods are not currently available.
When it comes to condensed matter systems, we are thus left with two approaches: model them with Dirac Hamiltonians~\cite{Abanov2001-p2, Ran2011, Grover2008, Yue2023, Yue2025} or consider systems where gradient expansions are adequate~\cite{Volovik1989, Moon1995, Iordanskii1997, Apel1997, Ray1999, Sengupta2000, Yakovenko1990, Yakovenko1991, Hlousek1990}.
Both have limitations.
Topological properties, like the value of $\theta$, are very often sensitive to the global momentum structure.
Tuning the chemical potential close to a Dirac point does not, therefore, guarantee that a Dirac Hamiltonian is appropriate for calculating $\theta$.
Gradient expansions, on the other hand, are incapable of finding the Hopf term when such common symmetries as reflection or time reversal are present.

\subsection{Lattice model realizations} \label{sec:lattice-model}
To enable $\vu{n}, \vu{n}'$ to fluctuate freely, we assume no external magnetic fields.
Thus TR symmetry is preserved on the level of the four-fermion interaction~\eqref{eq:f-interaction}, which requires that $\vu{n}, \vu{n}'$ (and their conjugates $\vb{S}, \vb{S}'$) be either even or odd under TR.
As we are interested in a radiatively-induced Hopf term, we shall assume that $\vu{n}$ and $\vu{n}'$ are locally aligned and of opposite TR signs.

From Eq.~\eqref{eq:Chern} we see that the Chern number $C$ vanishes identically when there is only one band because in that case the $X_{\mu}$ of Eq.~\eqref{eq:Xmua-def} are commuting numbers.
For a minimal model, we therefore need at least two orbital degrees of freedom per lattice site.

\begin{figure}[t]
\includegraphics[width=0.95\columnwidth]{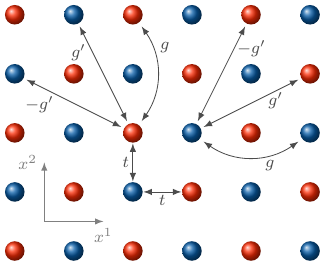}
\caption{A square lattice with $\vb{Q} = (\pi, \pi)$ antiferromagnetic ordering.
The $t$ arrows indicate the hoppings of the kinetic part of the Hamiltonian in Eq.~\eqref{eq:AF-Haml}.
The $g$ ($g'$) arrows indicate the momentum dependence of the coupling to $\vu{n}$ ($\vu{n}'$).
After ordering, the induced hoppings related to $g$ and $g'$ acquire additional $\pm$ signs that alternate with the colors of the sites.}
\label{fig:AF-lattice}
\end{figure}

As a first inter-valley magnetic example, let us consider a square lattice system that has two valleys, one electron-like and centered at the origin and another hole-like and centered at $\vb{k}_Z = (\pi, \pi)$.
For simplicity, we take that the dispersions of the two bands have the same form, but with opposite signs.
The minimal model describing this system is then given by
\begin{align}
G_k^{-1} &= \iu k_0 \one + \varepsilon_{\vb{k}} \uptau^3 + \gamma_{\vb{k}} \uptau^1 \Pauli^3 + \gamma_{\vb{k}}' \uptau^2 \Pauli^3,
\end{align}
where $\uptau^a$ are valley Pauli matrices, we have set $\vu{n} = \vu{n}' = \vu{e}_3$, giving the spin Pauli matrix $\Pauli^3$ above,
\begin{align}
\varepsilon_{\vb{k}} &= - \mu - 2 t \mleft(\cos k_1 + \cos k_2\mright),
\end{align}
and, for instance,
\begin{align}
\gamma_{\vb{k}} &= g \mleft[\sin(2k_1-k_2) + \sin(2k_2+k_1)\mright], \\
\gamma_{\vb{k}}' &= g' \mleft[\sin(2k_1+k_2) + \sin(2k_2-k_1)\mright].
\end{align}
Here $\mu$, $g$, and $g'$ are all real.
Let us set the overall energy scale by setting $t = 1$.
This model has parity symmetry with $M(P) = \uptau^3$ and $90^{\circ}$ rotation symmetry with $M(C_{\pi/2}) = \one$, but no reflection or TR symmetries.
Its Chern number equals:
\begin{align}
C &= \sgn(g g') \begin{cases}
- C(- \mu), & \text{when $\mu < 0$,} \\
6, & \text{when $0 < \mu < \sqrt{3}$,} \\
-2, & \text{when $\sqrt{3} < \mu < 2 \sqrt{2}$,} \\ 
2, & \text{when $2 \sqrt{2} < \mu < 4$,} \\ 
0, & \text{when $4 < \mu$.}
\end{cases}
\end{align}
We could have also chosen
\begin{align}
\gamma_{\vb{k}} &= g \sin k_1, \\
\gamma_{\vb{k}}' &= g' \sin k_2,
\end{align}
but at the expense of breaking the $90^{\circ}$ rotation symmetry.
The Chern numbers would then have equaled
\begin{align}
C &= \sgn(g g') \begin{cases}
- 2 \sgn(\mu), & \text{when $\abs{\mu} < 4$,} \\
0, & \text{when $\abs{\mu} > 4$.}
\end{cases}
\end{align}

\begin{figure}[t]
\includegraphics[width=\columnwidth]{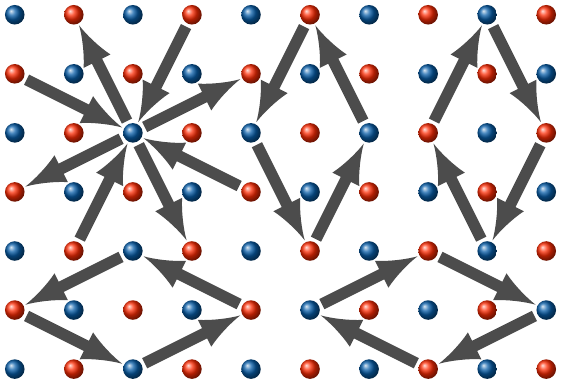}
\caption{The spin loop-current pattern that is induced by the coupling given in Eq.~\eqref{eq:LC-Haml}.
All spin currents emanating from one (blue) site are shown in the upper-left corner, from which we see that $90^{\circ}$ rotations are preserved, while reflections across $x^1$, $x^2$, and $x^1 \pm x^2$ are all violated by this pattern.
In light of Bloch's theorem~\cite{Bohm1949, Watanabe2019}, we can also depict this spin loop-current pattern in a more physical way as being made of circulating spin currents (right and lower parts of the figure).}
\label{fig:LC-pattern}
\end{figure}

In the antiferromagnetic case, the downfolding of the Brillouin zone effectively adds additional orbital degrees of freedom per unit cell.
A minimal antiferromagnetic model can therefore be constructed from only one orbital per lattice site.
Its hopping amplitudes are shown in Fig.~\ref{fig:AF-lattice} and the ordered-state Green function is given by
\begin{align}
\tilde{G}_k^{-1} &= \iu k_0 + \varepsilon_{\vb{k}} \uptau^1 + \gamma_{\vb{k}} \uptau^3 \Pauli^3 + \gamma_{\vb{k}}' \uptau^2 \Pauli^3, \label{eq:AF-Haml}
\end{align}
where $\vu{n} = \vu{n}' = \vu{e}_3$, giving the $\Pauli^3$ above, and
\begin{align}
\varepsilon_{\vb{k}} &= - 4 t \cos\tfrac{1}{2} k_1 \cos\tfrac{1}{2} k_2, \\
\gamma_{\vb{k}} &= 4 g \cos k_1 \cos k_2, \\
\gamma_{\vb{k}}' &= 8 g' \mleft(\cos k_1 - \cos k_2\mright) \sin\tfrac{1}{2} k_1 \sin\tfrac{1}{2} k_2. \label{eq:LC-Haml}
\end{align}
While $\gamma_{\vb{k}}$ (corresponding to $\vu{n}$) gives rise to a $\vb{Q} = (\pi, \pi)$ antiferromagnetic ordering, as depicted by the alternating red-blue colors in Fig.~\ref{fig:AF-lattice}, $\gamma_{\vb{k}}'$ ($\sim \vu{n}'$) induces the spin loop-current pattern shown in Fig.~\ref{fig:LC-pattern}.
This spin loop-current pattern alternates with the red-blue colors and respects four-fold rotation symmetry, but violates reflection symmetries across the principal and diagonal directions.
The chemical potential has been tuned to half-filling so that the Fermi surface has a $\vb{Q} = (\pi, \pi)$ nesting vector and becomes gapped upon ordering.
Let us set the energy scale with $t = 1$.
$g$ and $g'$ are real coupling constants.

If $\vu{e}_1$ and $\vu{e}_2$ are the initial primitive lattice vectors, then after ordering they become $\vb{E}_1 = \vu{e}_1 + \vu{e}_2$ and $\vb{E}_2 = \vu{e}_2 - \vu{e}_1$ (Fig.~\ref{fig:AF-lattice}).
The $k_1$ and $k_2$ are defined through $\vb{k} = \tfrac{1}{2} \mleft(k_1 \vb{E}_1 + k_2 \vb{E}_2\mright)$ and the fermionic field is made of $\mleft(\psi(\vb{R}), \psi(\vb{R}+\vu{e}_2)\mright)^{\intercal}$, where $\vb{R}$ go over the blue sites.
To make $\tilde{G}_k^{-1}$ periodic in momentum, we need to enact a gauge transformation
\begin{align}
G_k^{-1} = U_{\vb{k}}^{\dag} \tilde{G}_k^{-1} U_{\vb{k}},
\end{align}
where
\begin{align}
U_{\vb{k}} = \begin{pmatrix}
1 & 0 \\
0 & \Elr^{- \iu (k_1 + k_2) / 2}
\end{pmatrix} \Pauli^0.
\end{align}
This model has parity and $90^{\circ}$ rotation symmetry, but no reflection or TR symmetry due to $\gamma_{\vb{k}}'$.
Its Chern number is finite:
\begin{align}
C &= 4 \sgn(g g').
\end{align}
An alternative choice of spin coupling that breaks the $90^{\circ}$ rotation symmetry is
\begin{align}
\gamma_{\vb{k}} &= 2 g \cos k_1, \\
\gamma_{\vb{k}}' &= 2 g' \cos\tfrac{1}{2} (k_1 - k_2)
\end{align}
and it yields
\begin{align}
C &= - 2 \sgn(g g').
\end{align}

For both models, a finite Chern number $C$ implies that not only is the Hopf $\theta = - \pi C$ finite, as follows from Eq.~\eqref{eq:theta-Chern}, but also that there are gapless edge modes in the ordered state, which is a topological insulator belonging to the Altland-Zirnbauer Class~A~\cite{Chiu2016}.
Whether the system becomes ordered at $T = 0$, giving a topological insulator, or disordered, giving a bosonic Hopf SPT, depends on the magnitude of the phase stiffness $K$, which is defined in Eq.~\eqref{eq:NLSM-kinetic-part}.
There are contributions to $K$ coming from the initial interaction and from the electrons and one cannot, in general, say in what regime the system ends up.
However, as we discussed in detail in Sec.~\ref{sec:Hopf-properties}, for the Hopf SPT phase we know that it is gapped in the bulk, that it supports chiral gapless edge modes that carry spin preferentially in one direction, and that it exhibits a spin quantum Hall effect.

\section{Conclusion}
In this article, we have studied the properties and implications of a Hopf term and proposed the intersection of magnetic and spin loop-current order as a platform for realizing Hopf physics.
We have presented the proof of quantization of Hopf's theta in an accessible way and discussed the spin quantum Hall effect and chiral edge states that follow from a Hopf term.
After that, we introduced a very general model of a spin-ordered insulator, derived the Hopf term in this model using a gradient expansion, analyzed the conditions necessary for the appearance of a Hopf term, before finally exemplifying our proposal on a few lattice models.

The proposed platform for realizing a Hopf SPT requires simultaneous magnetic and spin loop-current order in a 2D system.
In recent years, there has been significant progress in identifying and engineering 2D magnetic materials~\cite{Gibertini2019, Huang2020, Blei2021, Petrovic2025}, which can be probed using magneto-optical Kerr effect microscopy, scanning magnetic imaging, or transport measurements.
Less forthcoming are 2D materials with spin loop-current order, in part because of the challenges in sensing such a ``hidden order'' that does not break time-reversal symmetry~\cite{Schulz1989, Ikeda1998, Zhou2017, Kontani2021, Shinjo2023}.
Second harmonic generation~\cite{Werake2010} and x-ray magnetic circular dichroism~\cite{Li2016} have been employed to probe injected spin currents and they may prove to be useful for studying equilibrium spin loop-current states.
Yet, the challenge in experimentally differentiating spin loop-current order from other time-reversal-preserving bond orders remains.
Spin loop currents have been suggested in the context of cuprates~\cite{Kontani2021} and the iridate Mott insulator \ce{Sr2IrO4}~\cite{Zhou2017, Kontani2021}, both of which are quasi-2D compounds that might exhibit the spin loop-current order, if truly present, even for atomically-thin samples.
Apart from this, it is worth noting that a 2D spin loop-current material, once found, can be integrated in many ways within heterostructures, as needed for realizing our proposal.

One unexpected finding of our whole analysis is that gradient expansions, contrary to popular belief, are not fully adequate for calculating topological terms, such as the Hopf term.
As we found in Sec.~\ref{sec:Hopf-symmetry}, due to the perturbative nature of the gradient expansion, symmetries that would normally imply quantization of the Hopf term instead imply its absence.
Furthermore, when the gradient expansion is applied to Dirac fermions, one wrongly finds no Hopf term, even though non-perturbative methods do find it.
Thus it is possible that Hopf terms are present in more general circumstances than we propose and study, shown in Fig.~\ref{fig:phase-diagram}.
Yet no methods are currently available to reliably calculate Hopf terms for general lattice models.
This remains an outstanding problem.

\begin{acknowledgments}
I thank Jörg Schmalian and Taylor L.\ Hughes for useful discussions.
This work was supported by the Deutsche Forschungsgemeinschaft (DFG, German Research Foundation) -- TRR 288-422213477 Elasto-Q-Mat Project No.\ A07.
\end{acknowledgments}

\appendix

\section{Hopf invariant in terms of linking numbers} \label{sec:link-formula}
Here we prove Eq.~\eqref{eq:Hopf-linking} of the main text.

Let us first observe that if $\vu{n}(x)$ changes by $\var[1]{\vu{n}} = \var{x^1} \partial_1 \vu{n}$ and $\var[2]{\vu{n}} = \var{x^2} \partial_2 \vu{n}$ for small displacements along $x^1$ and $x^2$, respectively, then the area spanned by $\vu{n}(x)$ in this small rectangle is given by $\vu{n} \vdot \mleft(\var[1]{\vu{n}} \vcross \var[2]{\vu{n}}\mright)$.
Up to normalization, this is precisely the $f^0(x)$ introduced in Eq.~\eqref{eq:topo-f-def} of the main text.
This explains expression~\eqref{eq:skyrmion} for the skyrmion number: it measures how many times the 2-sphere $\mathcal{S}^2$ in which $\vu{n}$ lives is spanned as we move around the 2D surface $\Sigma$.

On the 2-sphere $\mathcal{S}^2$, with suitable normalization, the Dirac delta function identity takes the form
\begin{align}
\int_{\mathcal{S}^2} \frac{\dd{S_{\vu{n}}}}{4 \pi} \Dd(\vu{n} - \vu{n}_0) \cdot h(\vu{n}) &= h(\vu{n}_0),
\end{align}
where $\dd{S_{\vu{n}}} = \sin \theta \dd{\theta} \dd{\phi}$ is the spherical area, $\vu{n} = \mleft(\sin \theta \cos \phi, \sin \theta \sin \phi, \cos \theta\mright)$, and $h(\vu{n})$ is an arbitrary function.
If we have a function $\vu{n}(x)$ that maps some 2D surface $\Sigma$ onto $\mathcal{S}^2$, then we can use it to pullback the above identity from $\mathcal{S}^2$ to $\Sigma$.
The result is
\begin{align}
\int_{\Sigma} \dd{\vb{S}_x} \vdot \vb{f}(x) \Dd(\vu{n}(x) - \vu{n}_0) \cdot h(x) &= \sum_{\ell} s_{\ell} h(x_{\ell}),
\end{align}
where the sum goes over all $x_{\ell} \in \Sigma$ whose $\vu{n}(x_{\ell}) = \vu{n}_0$ with the sign $s_{\ell} \equiv \sgn\mleft[\dd{\vb{S}_{x_{\ell}}} \vdot \vb{f}(x_{\ell})\mright]$.
If we integrate over the whole 3D space instead, the identity becomes
\begin{align}
\int \dd[3]{x} \, \vb{f}(x) \Dd(\vu{n}(x) - \vu{n}_0) \cdot h(x) &= \sum_{\ell} \oint_{\gamma_{\ell}(\vu{n}_0)} \dd{\vb{\ell}} \, h(x_{\ell}),
\end{align}
where the sum goes over all loops $\gamma_{\ell}(\vu{n}_0)$ made of $x_{\ell}$ such that $\vu{n}(x_{\ell}) = \vu{n}_0$ and orientated so that $\dd{\vb{\ell}} \vdot \vb{f}(x_{\ell}) > 0$.

By inserting
\begin{align}
1 &= \int_{\mathcal{S}^2} \frac{\dd{S_{\vu{n}_0}}}{4 \pi} \Dd(\vu{n}(x) - \vu{n}_0) \int_{\mathcal{S}^2} \frac{\dd{S_{\vu{n}_0'}}}{4 \pi} \Dd(\vu{n}(x') - \vu{n}_0')
\end{align}
into Eq.~\eqref{eq:Hopf-non-loc} and exploiting the above identity, one obtains
\begin{align}
Q_{\text{Hopf}} &= \sum_{\ell \ell'} \int \frac{\dd{S_{\vu{n}_0}}}{4 \pi} \frac{\dd{S_{\vu{n}_0'}}}{4 \pi} \operatorname{link}\mleft[\gamma_{\ell}(\vu{n}_0), \gamma_{\ell'}(\vu{n}_0')\mright], \label{eq:Hopf-linking-proof}
\end{align}
where $\operatorname{link}[\gamma, \gamma'] \in \Z$ is the linking number that appeared here through Gauss's linking integral:
\begin{align}
\operatorname{link}[\gamma, \gamma'] &= \frac{1}{4 \pi} \oint_{\gamma} \oint_{\gamma'} \frac{(\vb{\ell} - \vb{\ell}') \vdot \mleft[\dd{\vb{\ell}} \vcross \dd{\vb{\ell}'}\mright]}{\abs{\vb{\ell} - \vb{\ell}'}^3}.
\end{align}
The physical interpretation of this formula is as follows.
If we take that a unit current $I = 1$ flows along $\gamma$, $\vb{j}_{\gamma}(\vb{x}) = \oint_{\gamma} \dd{\vb{\ell}} \Dd(\vb{x} - \vb{\ell})$, then Biot–Savart's law gives the magnetic field $\vb{B}_{\gamma}(\vb{x})$.
Applying this to Gauss's linking integral, we find that $\operatorname{link}[\gamma, \gamma'] = \oint_{\gamma'} \dd{\vb{\ell}'} \vdot \vb{B}_{\gamma}(\vb{\ell}')$.
Since $\grad \vcross \vb{B}_{\gamma} = \vb{j}_{\gamma}$, $\operatorname{link}[\gamma, \gamma'] = \int_{S'} \dd{\vb{S}'} \vdot \vb{j}_{\gamma}$ where $\partial S' = \gamma'$.
Hence $\operatorname{link}[\gamma, \gamma']$ measures the number of times $\gamma$ punctures through $S'$, which is the same as the number of times $\gamma'$ wraps around $\gamma$.

The linking number is a topological invariant and is therefore a piecewise-constant function of $\vu{n}_0$ and $\vu{n}_0'$.
For sufficiently smooth and regular field configurations $\vu{n}(x)$, $\operatorname{link}\mleft[\gamma_{\ell}(\vu{n}_0), \gamma_{\ell'}(\vu{n}_0')\mright]$ is singular only at isolated points, i.e., on a set of measure zero.
The $\vu{n}_0$ and $\vu{n}_0'$ surface integrals in Eq.~\eqref{eq:Hopf-linking-proof} may thus be eliminated.
The result is Eq.~\eqref{eq:Hopf-linking}, subject to the condition that $\vu{n}_0$ and $\vu{n}_0' \neq \vu{n}_0$ are regular.

\section{Properties of the $\mathrm{SU}(2)$ and $\C P^1$ parameterizations of unit-vector fields} \label{sec:U-properties}
A unit-vector field
\begin{align}
\begin{aligned}
\vu{n}(x) &= \mleft(\hat{n}_1, \hat{n}_2, \hat{n}_3\mright) \\
&= \mleft(\sin \theta \cos \phi, \sin \theta \sin \phi, \cos \theta\mright)
\end{aligned}
\end{align}
can be parameterized in many ways.

In the main text, we found it useful to use an $\SU(2)$ group-valued field $U(x)$ to parameterize $\vu{n}(x)$.
The two are related through Eq.~\eqref{eq:SU2-param-def}:
\begin{align}
\vu{n} \vdot \vb{\Pauli} &= U \Pauli^3 U^{-1}.
\end{align}
Explicitly,
\begin{align}
\begin{aligned}
U = \, &\cos\tfrac{1}{2} \theta \, \mleft(\one \cos \phi_1 - \iu \Pauli^{3} \sin \phi_1\mright) \\
&+ \sin\tfrac{1}{2} \theta \, \mleft(\iu \Pauli^{1} \sin \phi_2 - \iu \Pauli^{2} \cos \phi_2\mright)
\end{aligned}
\end{align}
corresponds to the spherical angles $\theta$ and $\phi = \phi_1 + \phi_2$.

Using $\tr \Pauli^a \Pauli^b \Pauli^c = 2 \iu \LCs^{abc}$, it is easy to show that
\begin{gather}
\vu{n} \vdot \mleft(\partial_{\mu} \vu{n} \vcross \partial_{\nu} \vu{n}\mright) = - \iu \tr \Pauli^3 \mleft[A_{\mu}, A_{\nu}\mright], \\
\vb{a} \vdot \vb{f} = - \frac{\iu}{24 \pi^2} \LCs^{\mu \nu \rho} \tr A_{\mu} A_{\nu} A_{\rho},
\end{gather}
where $A_{\mu} \defeq U^{-1} p_{\mu} U$ with $p_{\mu} \defeq - \iu \partial_{\mu}$.
When the first of these two relations is combined with the gauge-free condition~\eqref{eq:gauge-free}, we get Eq.~\eqref{eq:a-A3-rel}, as follows from Eqs.~\eqref{eq:topo-f-def} and~\eqref{eq:vec-pot-a}.
The second relation proves the correspondence between the Hopf invariant of Eq.~\eqref{eq:Hopf-CS} and the $\SU(2)$ winding number invariant of Eq.~\eqref{eq:Hopf-U}.

Under a gauge transformation $U \mapsto U \, \Elr^{\iu \chi \Pauli^3}$:
\begin{align}
\begin{aligned}
\begin{pmatrix}
A_{\mu}^1 \\ A_{\mu}^2
\end{pmatrix} &\mapsto \begin{pmatrix}
\cos 2 \chi & - \sin 2 \chi \\
\sin 2 \chi & \cos 2 \chi
\end{pmatrix} \begin{pmatrix}
A_{\mu}^1 \\ A_{\mu}^2
\end{pmatrix}, \\
A_{\mu}^3 &\mapsto A_{\mu}^3 + \partial_{\mu} \chi.
\end{aligned}
\end{align}

For the gradient expansion of Sec.~\ref{sec:Hopf-gradient}, we need
\begin{gather}
\ev{U^{-1} p_{\mu} p_{\nu} U}_{s} = A_{\mu} A_{\nu}, \\
\ev{U^{-1} p_{\mu} p_{\nu} p_{\rho} U}_{s} = A_{\mu} A_{\nu} A_{\rho} \\
+ \tfrac{1}{2} A_{\mu} (p_{\nu} A_{\rho}) - \tfrac{1}{2} (p_{\nu} A_{\mu}) A_{\rho} + \tfrac{1}{8} \mleft(p_{\mu} p_{\nu} A_{\rho} + p_{\rho} p_{\nu} A_{\mu}\mright) \notag \\
+ \tfrac{1}{8} \mleft[A_{\nu} (p_{\mu} A_{\rho}) - A_{\mu} (p_{\rho} A_{\nu}) + (p_{\mu} A_{\nu}) A_{\rho} - (p_{\rho} A_{\mu}) A_{\nu}\mright], \notag
\end{gather}
where the last term vanishes when contracted with a symmetric tensor.
Here $\ev{\cdots}_s$ is the symmetrization defined in Eq.~\eqref{eq:sym-def} of the next appendix.

Although in the main text we did not use the $\C P^1$ parameterization, it is often employed in this context~\cite{Wu1984}.
Elements of the complex projective line $\C P^1$ are given by normalized pairs $\vb{z} = (z_0, z_1) \in \C^2$, $\vb{z}^{\dag} \vb{z} = 1$, where $\vb{z}$ and $\Elr^{\iu \chi} \vb{z}$ are identified for all phases $\Elr^{\iu \chi} \in \Ugp(1)$.
$\vu{n}$ in terms of $\vb{z}$ is given by the Hopf map:
\begin{align}
\vu{n} &= \begin{pmatrix}
z_0 z_1^* + z_0^* z_1 \\
\iu (z_0 z_1^* - z_0^* z_1) \\
\abs{z}_0^2 - \abs{z}_1^2
\end{pmatrix}.
\end{align}
$\vb{z} = \mleft(\cos\tfrac{1}{2} \theta \, \Elr^{- \iu \phi_1}, \sin\tfrac{1}{2} \theta \, \Elr^{\iu \phi_2}\mright)$ corresponds to the spherical angles $\theta$ and $\phi = \phi_1 + \phi_2$.
This is actually equivalent to the $\SU(2)$ group-valued field $U$:
\begin{align}
U &= \begin{pmatrix}
z_0 & - z_1^{*} \\
z_1 & z_0^{*}
\end{pmatrix}.
\end{align}
One may show that the $a_{\mu}$ of Eq.~\eqref{eq:vec-pot-a} is given in terms of $\vb{z}$ by:
\begin{align}
a_{\mu} &= - \frac{\iu}{2 \pi} \vb{z}^{\dag} \partial_{\mu} \vb{z}.
\end{align}

\section{Gradient expansion formula} \label{sec:gradient-formula}
Consider the one-particle operator:
\begin{align}
\mathcal{H} &= \sum_{\vb{R}_1 \vb{R}_2} \bar{\psi}(\vb{R}_1) U^{\dag}(\vb{R}_1) H(\vb{R}_1 - \vb{R}_2) U(\vb{R}_2) \psi(\vb{R}_2),
\end{align}
where $U(\vb{R})$ is a unitary matrix that commutes with the Hermitian matrix $H(\vb{R}) = H^{\dag}(-\vb{R})$.
If $U(\vb{R})$ varies slowly compared to the lattice constant, one may expand in its gradients.
This gradient expansion is given by
\begin{align}
\mathcal{H} &= \sum_{\vb{k}, \vb{q}} \sum_{n=0}^{\infty} \bar{\psi}_{\vb{k}+\vb{q}/2} O_{\vb{k}, \vb{q}}^{(n)} \psi_{\vb{k}-\vb{q}/2},
\end{align}
where
\begin{align}
O_{\vb{k}, \vb{q}}^{(n)} &= \frac{1}{n!} \sum_{i_1 \cdots i_n} \frac{\partial^n H_{\vb{k}}}{\partial k_{i_1} \cdots \partial k_{i_n}} \ev{U^{\dag} \ev{p_{i_1} \cdots p_{i_n}}_s U}_{\vb{q}}.
\end{align}
The Fourier transform conventions are
\begin{align}
\psi(\vb{R}) &= \mathcal{N}^{-1/2} \sum_{\vb{k}} \Elr^{\iu \vb{k} \vdot \vb{R}} \psi_{\vb{k}}, \\
H(\vb{R}) &= \mathcal{N}^{-1} \sum_{\vb{k}} \Elr^{\iu \vb{k} \vdot \vb{R}} H_{\vb{k}},
\end{align}
where $\mathcal{N} = \sum_{\vb{R}} 1$ is the number of unit cells and
\begin{align}
\ev{\cdots}_{\vb{q}} &= \mathcal{N}^{-1} \sum_{\bar{\vb{R}}} \Elr^{- \iu \vb{q} \vdot \bar{\vb{R}}} \big(\cdots\big)_{\bar{\vb{R}}}
\end{align}
is an averaged Fourier transform that is carried out with respect to $\bar{\vb{R}} \equiv \tfrac{1}{2} \mleft(\vb{R}_1 + \vb{R}_2\mright)$.
$\bar{\vb{R}}$ goes over a lattice that is $2^d$ times denser than the lattice of $\vb{R}$ ($d$ is the spatial dimension), but for finite samples still has $\sum_{\bar{\vb{R}}} 1 = \mathcal{N}$ (at a fixed $\vb{R}_1$ or $\vb{R}_2$).
The symmetrization is defined according to
\begin{align}
\ev{p_{i_1} p_{i_2} \cdots p_{i_n}}_s &= \frac{1}{2^n} \sum_{\ell=0}^{n} \binom{n}{\ell} \loarrow{p}_{i_1} \cdots \loarrow{p}_{i_{\ell}} \roarrow{p}_{i_{\ell+1}} \cdots \roarrow{p}_{i_n}, \label{eq:sym-def}
\end{align}
where $p_i = - \iu \, \roarrow{\partial}_i = \iu \, \loarrow{\partial}_i$.
For slowly varying fields, one may take the continuum limit
\begin{align}
\mathcal{N}^{-1} \sum_{\bar{\vb{R}}} \longrightarrow \frac{1}{L^d} \int \dd{\vb{r}},
\end{align}
where $L^d$ is the total volume.
This lattice gradient expansion formula is derived by Taylor expanding $U^{\dag}(\vb{R}_1) U(\vb{R}_2)$ around $\bar{\vb{R}} = \tfrac{1}{2} \mleft(\vb{R}_1 + \vb{R}_2\mright)$.

\bibliography{REFS-Hopf}

\end{document}